\documentstyle[eqsecnum,prc,preprint,aps]{revtex}
\begin{document}
\tighten
\draft
\title{Nuclear muon capture by \mbox{$^{3}$He}: meson exchange currents for the
triton channel}
\author{J.G.Congleton\thanks{Supported by the Sir James Knott Fellowship.}}
\address
{Department of Physics, University of Newcastle, Newcastle upon Tyne, 
NE1 7RU, UK.}
\author{E.Truhl\'{\i}k\thanks{Supported by grants Nos.\ 
202/94/0370 (GA CR) and 148410 (GA ASC CR).}}
\address
{Inst.\ of Nuclear Physics, Academy of Sciences of the Czech Republic,\\
CZ 250 68 \v{R}e\v{z} near Prague, Czech Republic.}
\date{\today}
\maketitle
\begin{abstract}
We have calculated exchange corrections for nuclear muon capture by
\mbox{$^{3}$He} leading to the \mbox{$^{3}$H} final state using the hard-pion
model and realistic nuclear wavefunctions. These currents modify the
vector and axial part of the weak nuclear current. In their absence
the rate is 12\% smaller than found by experiment.

Our final result for the rate is $1502 \pm 32$ per second. For the
analyzing powers we find $A_v =0.515 \pm 0.005$, $A_t=-0.375 \pm
0.004$ and $A_\Delta = -0.110 \pm 0.006$. These predictions use the PCAC
value of $g_P$. The variation of the observables with $g_P$ is also
reported.

\end{abstract}
 
\narrowtext
\section{Introduction}

As has recently been discussed in Ref.~\cite{JCHF}, the reaction of
negative muon capture by the helion,
\begin{equation}\label{reac}
\mu^- + \mbox{$^{3}$He} \rightarrow \nu_\mu + \mbox{$^{3}$H}
\end{equation}
is at present potentially able to provide us with the induced
pseudoscalar coupling $g_P$ with nearly the same precision as from
capture by a free proton. Indeed, accurate three-nucleon
wavefunctions are now available and the uncertainties due to the
description of the nuclear states can be reduced to a minimum.

However, the calculations reported in Refs.~\cite{ACST,AHHST} for
other weak reactions on light nuclei show that besides the
one--nucleon contribution, meson exchange current (MEC) effects should
be taken into account.  That this is so can also be seen from the
analysis performed in Ref.~\cite{JCHF}, where Congleton and Fearing
compared the results obtained using the elementary particle model
(EPM) to the standard calculations in the impulse approximation (IA):
the effective magnetic couplings $G_{P}$ and $G_{A}$ are 10\% smaller
in the IA (see Table~\ref{IAG}). Table~\ref{IAj} shows the
contributions of the current components to the effective couplings,
and it can be seen that $\vec{j}_{\,A}$ makes an important
contribution to $G_{P}$ and $G_{A}$.

Here we continue the study of the characteristics of
reaction~(\ref{reac}) and consider the weak MEC effects. For the
operator of the weak axial nuclear MEC we adopt the one published
recently in Ref.~\cite{TK}, which was applied earlier to negative muon
capture by deuterons in Ref.~\cite{ACST}. The weak axial nuclear MEC
operator satisfies the nuclear continuity equation (PCAC) up to the
order considered, which is $1/M^{\,2}$ (M is the nucleon mass). The
space-like component of the operator contains both static and velocity
dependent parts. We take into account fully both the static part and
the terms linear in external momenta from the velocity dependent part.

Further, we use for the vector part of the weak MEC operator the
standard isovector currents well known from the electromagnetic
processes~\cite{LAmag,SHS,HB,CR,TA}. We employ non--relativistic
operators satisfying CVC at order $1/M$.

A brief account of the work has been recently reported
in Ref.~\cite{CT}. Here we present the full results. In
section~\ref{forma}, we discuss the formalism employed,
section~\ref{resul} contains the numerical analysis of the studied
problem and in section~\ref{concl} we give our conclusions. The most
important of them is that including the weak axial and vector
$\pi$-MEC in the microscopic calculation yields results which agree
closely with the EPM in their predictions for observables.

In order to make the paper more transparent, we postpone necessary
details of the formalism and partial results into a series of
appendices.

\section{Formalism}
\label{forma}

To evaluate the effect of meson exchange currents, the $\mbox{$^{3}$H}
\rightarrow \mbox{$^{3}$He}$ weak nuclear current was parameterized by six $Q$
dependent current amplitudes
$\rho_0,\rho_1,j_Q,j_\sigma,j_\times^{(1)}$ and $j_\times^{(2)}$ as
follows.
{\mathletters
\label{c1}
\begin{eqnarray}
\langle \, \mbox{$^{3}$H} ; m_f\mid j^0 \mid \mbox{$^{3}$He} ; m_i 
\,\rangle &=&
\chi_{m_f}^{\dag} \Bigl\{ \rho_0 + \rho_1 \hat{Q}\cdot \vec{\sigma}
\Bigr\}
\chi_{m_i} \\
\langle \, \mbox{$^{3}$H} ; m_f\mid \vec{j} \mid \mbox{$^{3}$He} ; m_i 
\,\rangle &=&
\chi_{m_f}^{\dag} \Bigl\{ j_Q \hat{Q} 
+ j_\times^{(1)} i\hat{Q} \times \vec{\sigma}
+ j_\sigma \vec{\sigma} 
+ j_\times^{(2)} [\hat{Q} (\hat{Q} \cdot \vec{\sigma}) -
 \case{1}{3}\vec{\sigma}] \Bigr\} \chi_{m_i}
\nonumber \\
\end{eqnarray}}

In Eqs.\ (\ref{c1}), $m_i$ and $m_f$ are the initial and final
projections of the internal nuclear angular momentum, $j^0$ and
$\vec{j}$ are the time and space components of the total weak four
current and the three-momentum transfer to the helion is $\vec{Q}$.
The current amplitudes are not relativistically invariant and are not
independent (if we make the assumption that second class currents are
absent). To see the non-independence note that the EPM parameterizes
the current using only four form factors whereas there are six current
amplitudes. The current amplitudes are, however, useful in the
formulation because there exists a simple correspondence between the
non-relativistic current operators and the current amplitude to which
the matrix element of that operator contributes.
 
The total current has a vector and an axial part. The vector part
contributes to $\rho_0,j_Q,j_\times^{(1)}$ and the axial part to
$\rho_1,j_\sigma,j_\times^{(2)}$.  The relationships between these
current amplitudes and the traditional effective form factors
$G_V,G_P$ and $G_A$ are,
\begin{mathletters}
\label{eff}
\begin{eqnarray}
G_V &=& \rho_0 + j_Q \\
G_P &=& j_\times^{(1)} - j_\times^{(2)} - \rho_1 \\
G_A &=& j_\times^{(1)} + j_\sigma - \frac{1}{3}j_\times^{(2)}.
\end{eqnarray}
\end{mathletters}

The statistical capture rate $\Gamma_0$ and analyzing powers $A_v,A_t$
and $A_\Delta$ for reaction (\ref{reac}) can be found from the
effective form factors by using Eqs.\  (3)(11)(12) and (13)
of Ref.~\cite{JCHF}.

Our model for the nuclear current employs non-relativistic
wavefunctions and current operators the latter of which are listed in
appendix~\ref{appc}.  After suffering a multipole decomposition of the
plane wave factor, the current operators yielded matrix elements with
the following five forms,
{\mathletters
\widetext
\begin{equation}
Z_{l\lambda L}^J = \langle \, \mbox{$^{3}$H} \mid \mid
[ 4\pi {\cal Y}_{l\lambda}^L(\hat{x}\hat{y}) 
\otimes S_\Sigma]_J \: j_\lambda( Q y /3) {\cal I}
\mid \mid \mbox{$^{3}$He} \, \rangle / (2J+1)^{1/2} 
\end{equation}
\begin{equation}
_xZ_{l\lambda L}^J = \langle \, \mbox{$^{3}$H} \mid \mid
[ 4\pi {\cal Y}_{l\lambda}^L(\hat{x}\hat{y}) 
\otimes S_\Sigma]_J \: j_\lambda( Q y /3) {\cal I}
\frac{\partial}{\partial x_\pi}
\mid \mid \mbox{$^{3}$He} \, \rangle / (2J+1)^{1/2} 
\end{equation}
\begin{equation}
_yZ_{l\lambda L}^J = \langle \, \mbox{$^{3}$H} \mid \mid
[ 4\pi {\cal Y}_{l\lambda}^L(\hat{x}\hat{y}) 
\otimes S_\Sigma]_J \: j_\lambda( Q y /3) {\cal I}
\frac{\partial}{\partial y_\pi}
\mid \mid \mbox{$^{3}$He} \, \rangle / (2J+1)^{1/2} 
\end{equation}
\begin{equation}
_xZ_{l\lambda L}^{K1J} = \langle \, \mbox{$^{3}$H} \mid \mid
\Bigl[[ 4\pi {\cal Y}_{l\lambda}^L(\hat{x}\hat{y}) 
\otimes S_\Sigma]_K
\otimes \frac{\hat{\nabla}_x}{x_\pi}\Bigr]_J \: j_\lambda( Q y /3) {\cal I}
\mid \mid \mbox{$^{3}$He} \, \rangle / (2J+1)^{1/2} 
\end{equation}
\begin{equation}
_yZ_{l\lambda L}^{K1J} = \langle \, \mbox{$^{3}$H} \mid \mid
\Bigl[[ 4\pi {\cal Y}_{l\lambda}^L(\hat{x}\hat{y}) \otimes S_\Sigma]_K
\otimes \frac{\hat{\nabla}_y}{y_\pi}\Bigr]_J \: j_\lambda( Q y /3) {\cal I}
\mid \mid \mbox{$^{3}$He} \, \rangle / (2J+1)^{1/2} .
\end{equation}}
\narrowtext
The letter $Z$ takes on values $A \ldots I$ according to the spin and
isospin operators $S_\Sigma$ and ${\cal I}$ as indicated in
Table~\ref{AI}. For example, the Gamow-Teller matrix element is 1/4
$F_{000}^1$. The matrix elements are reduced in the spin-spatial part but not in
the isospin part so that for example,
{\mathletters
\begin{equation}
\langle \, \mbox{$^{3}$He} \mid \mid 3 \tau_3^z  
\mid \mid \mbox{$^{3}$He} \, \rangle = +1
\end{equation}
\begin{equation}
\langle \, \mbox{$^{3}$H} \mid \mid 3 \tau_3^z  
\mid \mid \mbox{$^{3}$H} \, \rangle = -1 .
\end{equation}}
The following selection rules were applied to the matrix
elements.
\begin{enumerate} 
\item Parity. For $Z_{l\lambda L}^J$ this implies that only matrix
elements with, $l+\lambda = \mbox{even}$, are significant.
\item Hermiticity and isospin symmetry. Because \mbox{$^{3}$He} and 
\mbox{$^{3}$H} form a good isospin doublet, matrix elements 
$Z_{l\lambda L}^J$ with an anti-hermitian operator are
insignificant. This approximation is good at the few $\times 10^{-3}$
level since we have for the wavefunctions used here, (which have no
isospin 3/2 components),
\begin{equation} \label{isosym}
\langle \, \mbox{$^{3}$H} \mid \mid 3 I_3^- \mid \mid \mbox{$^{3}$He} 
\, \rangle = 0.9998 \end{equation} 
and further the probability of the $I=3/2$ components is
$10^{-5}-10^{-6}$~\cite{Sasa}.  For the local matrix elements
$A,C,D,E,F$ this implies that only $J=1$ is significant and for
$B,G,H,I$ only $J=0$ is significant. A numerical example is
$E_{110}^0/(\surd{3} E_{110}^1) = 9\times 10^{-4}$. \item Selection
rule for $l$. If the spin-isospin part of the operator is even(odd)
under the interchange or particle labels 2 and 3 then only matrix
elements for which $l$ is even(odd) are significant e.g.\ $A$ requires
$l=\mbox{odd}$. This selection rule follows from the symmetry of the
wavefunctions under interchange of particle labels.
\item A selection rule peculiar to the $F$ type matrix elements is
that $F_{111}^1$ is zero by interchange of particle labels 1 and 2.
\end{enumerate}

Further to the selection rules a rationalization of the matrix
elements was possible by taking into account the long wavelength of
the $W^{\pm}$ boson mediating the interaction, the low P-state
probability ($<0.2\%$) and the $\sim9\%$ D-state probability in the
trion wavefunctions. With regard to the first point we note that the
isovector spectator point-particle range is about 1.7 fm. This implies
that matrix elements with high values of $\lambda$ are small and that
a power expansion in $y$ rapidly converges, e.g.\ the $y^2$ term in
$j_0(Qy/3)$ contributes at the 3\% level. The range of the pair
coordinate $x$ is of the order of the pion Compton wavelength, 1.4 fm,
as can be seen from Fig.~\ref{densities}.  This is expected because
the exchange current operators pick out the spin and isospin
dependence in the wavefunction which is mostly due to pion
exchange. Matrix elements with the factor $j_f( Q x/2)$ will be small
for large values of $f$.  

Integrals of the densities are given in Table~\ref{MEint}. An example
of rationalization is to consider the ratios $D_{202}^1/D_{000}^1 =
2\times 10^{-1}$ and $D_{022}^1/D_{202}^1=1\times10^{-2}$.  We note
that $D_{000}^1$ receives contributions solely from SS
overlap. $D_{202}^1$ and $D_{022}^1$ are dominated by SD overlap and
are thus diminished due to the lower D-state probability by a factor
$\sim\surd(0.09/0.91) = 3\times 10^{-1}$ which explains the ratio of
$D_{202}^1$ and $D_{000}^1$. $D_{022}^1$ is further suppressed due to
the low momentum transfer in the process by the $j_2(Qy/3)$
factor\footnote{Note that one-body currents have a $j_\lambda(2Qy/3)$
factor showing that the effective momentum transfer for exchange
currents is half as much as for single-nucleon currents.} $\sim 3
\times 10^{-2}$. This explains why $D_{022}^1$ is so much smaller than
$D_{202}^1$.

For most local currents the leading multipoles plus those within one
unit of angular momentum were included. This procedure neglects
contributions at the 3\% level. For the largest current (delta
excitation) these 3\% corrections were included.

The matrix elements $_y Z_{l\lambda L}^J$ and $_y Z_{l\lambda
L}^{K1J}$ were neglected since we expect their contributions to be
small compared to those of $_xZ_{l\lambda L}^J$ and $_x Z_{l\lambda
L}^{K1J}$. They will be suppressed by about 15\% due to angular
momentum propensity rules. By parity we need an extra factor $\vec{x}$
or $\vec{y}$ to combine with $\nabla_y$. In the case of $\vec{x}$ this
yields $l=\lambda = 1$ and these elements are suppressed as can be
seen in Table~\ref{MEint}.  In the case of $\vec{y}$ there will be
suppression because of the $j_1(Qy/3)$ factor of about 15\%.

The fact that we omit the non-local terms involving $\nabla_y$ implies
that our estimate of the non-local terms has an inherent uncertainty
of about 10-20\%. It turns out that the non-local currents due to
$\nabla_x$ contribute at the 10\% level to the exchange currents and
so the above neglect affects the results for the MEC at the 1-2\%
level and the results for the total current at less than the 0.5\%
level.

Each exchange current yields a contribution to one, (or more), of the
current amplitudes and the general form is,
\begin{equation}
j = \int dx \, \rho (x) \, f(x)
\end{equation}
where $\rho(x)$ is a nuclear density and $f(x)$ is a `potential
function' which depends on the meson which is exchanged, the overall
coupling strength and also the momentum dependence of the current.
The definitions of the various potential functions entering the
calculation are given in appendix~\ref{apppot}.

The nuclear densities were calculated using wavefunctions found by the
coupled rearrangement channel (CRC) method of Kameyama {\em et
al.}~\cite{Kami}. The Faddeev components are expressed as sums of
gaussians and one can find analytic expressions for the matrix element
densities as functions of $x$, the pair coordinate. The densities were
evaluated at either 14 points or in the case of the large matrix
elements (first five of Table~\ref{MEint} and the matrix elements with
a derivative i.e.\ $_xZ_{l\lambda L}^J$), at 24 points with a higher
density of points for $x<1.5\mbox{ fm}$. The reason for calculating a
density is to facilitate the calculation of matrix elements with
different potential functions which necessity arises when meson
coupling parameters and strong form factor cutoffs
$\Lambda_\pi,\Lambda_\rho$ and $\Lambda_a$ are varied.

The calculation of the local matrix elements was checked by comparing
results for the trion
%
%
isovector magnetic moment, $\mu_v$, with those of Friar {\em et al.}
\cite{LAmag}. In that work, wavefunctions for the triton were applied
resulting from many NN potentials one of which was the AV14 potential:
only exchange currents arising from $\pi$ exchange were
considered. With $\pi$ exchange only, the expression for $\mu_v$ in
our formalism is, {\mathletters
\begin{eqnarray}
\mu_v &=& \lim_{Q \rightarrow 0} (-)\frac{m_p}{Q} j_x^{(1)}
\hspace{2ex}\mbox{ n.m.}  \\
&=& \mu_v^{\rm IA} +  \mu_v^{\rm MEC}
\end{eqnarray}
where,
\begin{eqnarray}
\mu_v^{\rm IA} &=& \case{1}{2}  ( 1 + \kappa^p - \kappa^n )
[{\bbox \sigma }]^{0,1} \\
\mu_v^{\rm MEC} &=& \mu_v^{\rm pair} + \mu_v^\pi + \mu_v^{\Delta}
\end{eqnarray}
and,
\widetext
\begin{eqnarray}
\mu_v^{\rm pair} = (-) \frac{m_p}{m_\pi} f_{\pi NN}^2 \Bigl \{ 
&&\case{1}{\surd 2}   D_{000}^1[\case{x_\pi}{6} f_1]
+ \case{1}{2}       D_{202}^1[\case{x_\pi}{6} f_1]
+ \case{1}{\surd 6} C_{111}^1[\case{y_\pi}{9} f_1]
\nonumber \\ &&
-\case{5}{\surd 24} E_{111}^1[\case{y_\pi}{9} f_1]
- \case{5}{\surd 8} E_{110}^1[\case{y_\pi}{9} f_1] \Bigl \}
\end{eqnarray}
\begin{eqnarray}\label{muvpi}
\mu_v^{\pi} &=& (-) \frac{m_p}{m_\pi} f_{\pi NN}^2 \Bigl \{ 
\case{1}{\surd 2}   D_{000}^1[\case{1}{3} d_3 - d_2]
+ \case{1}{6}       D_{202}^1[d_3]   
+ \case{1}{\surd 6} C_{111}^1[\case{y_\pi}{9} ( d_5 - 5 d_3 / x_\pi)] 
\nonumber \\
&&-\case{1}{\surd 30} E_{111}^1[\case{y_\pi}{9} ( d_5 - 5 d_3 / x_\pi)]
- \case{1}{\surd 10} E_{112}^1[\case{y_\pi}{9} ( d_5 - 5 d_3 / x_\pi)]
\nonumber \\
&&- \case{1}{\surd 15} E_{312}^1[\case{y_\pi}{9} d_5 ]
- \sqrt{\case{2}{15}} E_{313}^1[\case{y_\pi}{9} d_5 ] \Bigl \}
\end{eqnarray}
\begin{eqnarray}
\mu_v^\Delta = 2 G_1 f_{\pi N\Delta} f_{\pi NN} \frac{4m_\pi m_p}
{9M(M_\Delta-M)} \Bigl \{ 
&& F_{000}^1[f_3 - f_2/2] +
\case{\surd 2}{3} F_{202}^1[f_2]   
\nonumber \\ &&
+\case{\surd 2}{4} D_{000}^1[f_3 - f_2/3] 
- \case{1}{12} D_{202}^1[f_2] \Bigr \}
\end{eqnarray}}
\narrowtext
where $\kappa_p$ and $\kappa_n$ are the proton and neutron anomalous
magnetic moments and $\kappa_p - \kappa_n = 3.706$.

The calculation of Friar {\em et al.} \cite{LAmag} used a different
potential function for $\mu_v^\pi$ arising from the propagator
$\Delta_F^\pi(Q_2)\Delta_F^\pi(Q_3)F_{\pi NN}(Q_2) F_{\pi NN}(Q_3)$
which, although an intuitive choice, is inconsistent with the equation
of continuity. To facilitate comparison we must replace $d_2
\rightarrow f_{25}, d_3 \rightarrow f_{26}$ and $d_5 \rightarrow
f_{27}$ in Eq.\ (\ref{muvpi}) (see appendix~\ref{apppot} for the
definitions of functions $f_i$) where $f_{25},f_{26},f_{27}$
correspond to the choice of propagator made in Ref.~\cite{LAmag}. For
the $\Delta$-excitation graph we need to choose $f_{\pi
N\Delta}=6\surd 2/5 f_{\pi NN}$ and $G_1=3\surd 2 /10
(\mu_p-\mu_n)=2.00$ so that our coefficients agree.

The comparison is shown in Table~\ref{3Nmag}.  Our results for the
exchange contributions are in overall agreement (case (b)) although
the comparison is not exact because we have used a triton-helion
overlap.  Also shown are results with \mbox{$^{3}$H} bra and ket wave
vectors and $\tau_3^-$ replaced with $-\tau_3^z$ (case (a)) which
corresponds exactly to the calculation of~\cite{LAmag}. These results
agree well.  The agreement gives us confidence in our calculation of
local matrix elements.  We also give results for 8 and 22 channel
wavefunctions from the AV14 potential with the Tuscon-Melbourne
three-body force (cases (c) and (d)) which show the effect of adding
extra channels to the Faddeev wavefunction. The effect is very
small. Although a direct comparison of our results with~\cite{LAmag}
is not possible because a different coupling scheme was employed there
(our wavefunctions employ Russell-Saunders LS coupling rather than
$jj$) our 8 and their 5 channel wavefunctions have similar content and
convergence of the binding energy requires 22 and 34 channels
respectively. The extra channels change the pair, pion and delta
contributions by less than 1\% in our case whereas the difference
in~\cite{LAmag} is about 9\%. This is a manifestation of the faster
convergence obtained when using the CRC method: the projection of the
potential onto partial waves is nearer to being complete for a given
number of Faddeev component channels for the CRC method as compared to
the method used in Ref.~\cite{LAmag}.

To check the calculation of the non-local matrix elements we used a
Peterson type device~\cite{Pete}. The device follows from a simple
identity for the matrix element of an operator ${\cal O}$ multiplied
by the momentum operator $\hat{p}$ when the bra and ket vectors are
equal. If,
\begin{equation}
M = \langle \psi \mid {\cal O} \hat{p} \mid \psi \rangle
\end{equation}
then for ${\cal O}$ hermitian $({\cal O}^\dagger = {\cal O})$,
\begin{equation}
\Im M = \frac{1}{2i} \langle \psi \mid [{\cal O},\hat{p}] \mid \psi \rangle
\end{equation}
and for ${\cal O}$ anti-hermitian $({\cal O}^\dagger = - {\cal O})$,
\begin{equation}
\Re M = \frac{1}{2} \langle \psi \mid [{\cal O},\hat{p}] \mid \psi \rangle.
\end{equation}

Applying the above to the operator ${\cal O}= i\sigma_3 \cdot \vec{x}
\, \tau_3$ we obtained,
\begin{equation}
\label{PT}
\case{3}{2}\; 
^{tt}F_{000}^1 = -_x ^{tt}F_{000}^1 + \surd 2 \; ^{tt}_xF_{202}^1 + 
3 \; _x^{tt}F_{101}^{011} .
\end{equation}
The `tt' superscript indicates that the triton spin-space
wavefunctions were used in both bra and ket and that $Q=0$. Eq.\
(\ref{PT}) relates a local matrix element $F_{000}^1$ to non-local
matrix elements both {\em with} $(_xF_{000}^1,_xF_{202}^1)$ and {\em
without} $(_xF_{101}^{011})$ derivatives. The relation~(\ref{PT}) was
found to hold to within the precision\footnote{The expansion
coefficients for the basis functions are know to a finite
precision. By noting the largest contribution to the matrix element
one can calculate the precision of the matrix element.}
of the calculation and to 0.4\% when the ket was replaced by a
\mbox{$^{3}$He} wavefunction. This deviation is first order in small
differences between the \mbox{$^{3}$He} and \mbox{$^{3}$H}
wavefunctions whereas the deviation given in equation (\ref{isosym})
is second order which explains why the latter is smaller.

A further check of the calculation was made by calculating the
Gamow-Teller matrix element, M(GT), which measures the axial current
at zero momentum transfer and governs the $\beta$-decay of the
triton. Our results are compared to those of Adam {\em et al.} \cite{AHHST} in
Table~\ref{GT}. The calculation reported in Ref.~\cite{AHHST} used
axial currents which, except for the $\Delta$-excitation currents, are
equal to those used here and applied various wavefunctions and meson
parameters (the values given in the table are for the Paris
potential~\cite{Paris}). For the purposes of Table~\ref{GT} we have
adopted the meson parameters given in Eq.\ (3.2) of
Ref.~\cite{AHHST}. However, because our hadronic form factors are
monopole $F(Q^2)=[(\Lambda^2-m^2)/(\Lambda^2+Q^2)]^n$ with $n=1$ and
those used by Adam {\em et al.} had $n=\case{1}{2}$, we applied
equivalent values for $\Lambda$ found by equating the slope of
$F(Q^2)$ with respect to $Q^2$ at the on shell point $Q^2 =
-m^2$. This procedure yielded, {\mathletters
\begin{equation} 
\Lambda_\pi (\mbox{monopole}) = 1.69 \mbox{ GeV}
\leftrightarrow \Lambda_\pi (n=\case{1}{2}) = 1.2 \mbox{ GeV}
\end{equation}
\begin{equation}
\Lambda_\rho (\mbox{monopole}) = 2.72 \mbox{ GeV}
\leftrightarrow \Lambda_\rho (n=\case{1}{2}) = 2.0 \mbox{ GeV.}
\end{equation}}
Taking into account the different wavefunctions used the results agree
quite well.

To check the calculation of the $\Delta$-excitation currents we
compared our results for M(GT) to those of Carlson {\em et al.}
\cite{CRSW}. That calculation used wavefunctions found from the AV14
NN potential and the Urbana three-body force. Our wavefunction should
therefore be only slightly different since the three-body force has
only a small effect.  To make the coupling coefficients the same we
set $f_{\pi NN}^2/4\pi = 0.079$, $f_{\pi \Delta N}^2/4\pi = 0.2275$,
$g_A(0)=-1.262$, $g_{\rho NN}^2/4\pi = 0.5$, $\kappa_V=6.6$ and
$G_1=3.06$. The pion and rho cutoffs were set to $\Lambda_\pi = 0.90
\mbox{ GeV}$, $\Lambda_\rho = 1.35 \mbox{ GeV}$ which makes our
potential functions agree exactly since Carlson {\em et al.} used
monopole form factors.  With these parameters we found that M(GT)
received contributions +0.052 and --0.022 from the currents ANP4 and ANP5
respectively which agree well with the values reported in Table I of
Ref.\cite{CRSW}.

\section{Results}
\label{resul}
We precede the presentation of results with a discussion of our
choices of parameters.  The ranges used for the parameters which
define the exchange currents are listed in Table \ref{param}. Our aim
is to make a realistic estimate of the theoretical uncertainty and so
we choose large ranges of reasonable values. 

We took the ratio of the rho anomalous to normal coupling,
$\kappa_{\rm V}$, to vary between 3.7 and 6.6. In the context of the
hard-pion model, which combines vector meson dominance (VMD) with
chiral symmetry, the value 3.7 is consistent as this is the VMD
value. However, the Bonn meson exchange force model OBEPR \cite{Bonn}
requires $\kappa_V = 6.6$. A similar criterion was applied to find
the range of the $\rho NN$ coupling where the hard-pion model requires
the KSRF \cite{KSRF} value $g_{\rho NN}^2/4\pi = 0.70$ whereas OBEPR
favours a stronger coupling of $g_{\rho NN}^2/4\pi = 0.95$. The value
of the $\rho N\Delta$ coupling, $G_1$, is taken from experimental data
for the M1/E2 multipole ratio for photoproductions of pions
\cite{Davi}. 
The parameters $\nu$ and $\eta$ reflect the off-shell ambiguity in the
pion exchange potential. For consistency with the AV14 potential,
which is a static potential, their values should be taken to be 1/2.
Our exchange current APSPV expresses the difference between what would
be obtained using pseudoscalar and pseudovector $\pi NN$
coupling. Pseudoscalar coupling corresponds to $\lambda=1$ and
pseudovector coupling to $\lambda=0$. It is not possible to achieve
exact consistency of this current with the NN potential but the value
$\lambda=1/2$ is the most appropriate. We varied $\lambda$ between 0
and 1.  
The value of $f_{\pi NN}^2/4\pi$ should be taken as 0.081 to
be consistent with the AV14 potential. However, the range 0.075 -
0.081 was used to estimate the uncertainty due to $f_{\pi NN}$ where
the low value comes from a recent Nijmegen analysis of NN scattering
data \cite{Nij93}.

The most influential parameters are the strong form factor cutoffs
$\Lambda_\pi,\Lambda_\rho,\Lambda_a$ and the $\pi N\Delta$ coupling
$f_{\pi N\Delta}$.  The strong form factors affect the potential
functions $f_i(x)$ entering the reduced matrix elements and in general
reduce their strength at small $x$ i.e.\ $x < \sim 1/\Lambda$. The
value $\Lambda_M = m_M$ where $M=\pi, \rho$ or $a_1$ corresponds to
complete cutoff. For example the current for $\Delta$-excitation due
to $\rho$ exchange, ANP5, would be zero if the choice $\Lambda_\rho =
m_\rho$ was made. We used the range $\Lambda_\pi = 1.0 - 1.5 \mbox{
GeV}$ which corresponds to a central value of 1.2~GeV with a 20\%
variation in $1/\Lambda_\pi$ either side.  The values of
$\Lambda_\rho$ and $\Lambda_a$ were then found by demanding consistency
between the exchange current and the NN potential used to construct
the wavefunction. The exchanges of $\pi,\rho,a_1$ yield isospin
dependent spin-spin and tensor interactions ($V_S$ and $V_T$) and it
is these two components of the AV14 NN force which were fitted. The
long range parts of $V_S$ and $V_T$ are fitted exactly provided the
same pion coupling $f_{\pi NN}$ and pion mass are used. The short
range part is affected by the strong form factors and the desired
effect of the form factors is that the strength of the potential at
short distance is reduced. Monopole form factors achieve this effect
for $V_T$ but for $V_S$ an undesirable change in sign at small
distance occurs. This can be traced back to the second derivative of
the potential function arising from $\pi,\rho$ or $a_1$ exchange. To
illustrate this consider the $\pi$ OBEP potential (see appendix
\ref{appOBEP} for further details).  {\mathletters
\begin{equation}
V_T^\pi \sim F_\pi^{''} - \frac{F_\pi^{'}}{x_\pi} \end{equation}
\begin{equation}
V_S^\pi \sim F_\pi^{''} + 2 \frac{F_\pi^{'}}{x_\pi} 
\end{equation}}
where $F^{'}= dF/dx_\pi$. Fig.\ \ref{figOBEP} shows
$V_T^\pi,V_S^\pi,F_\pi^{'}$ and $F_\pi^{''}$ for $\Lambda_\pi
\rightarrow \infty$ and $\Lambda_\pi = 1.2 \mbox{ GeV}$. The function
$F_\pi^{''}$ changes sign at $x_\pi \approx 0.16$ and we also have,
\begin{equation}
x_\pi F_\pi^{''} \rightarrow  F_\pi^{'} \mbox{ as } x \rightarrow 0.
\end{equation}

This explains why the combination of $F_\pi^{''}$ and $F_\pi^{'}$ in
$V_T$ does not change sign at small $x$ despite containing
$F_\pi^{''}$. The combination $(V_T - V_S)$ eliminates
$F_\pi^{''}$. In order not to be sensitive to the short range part of
$F_\pi^{''}$ we fitted $\Lambda_\rho,\Lambda_a$ to $V_T$ and
$(V_T-V_S)$ where the latter combination eliminates $F_\pi^{''}$. {\em
A posteriori} we noticed that the largest exchange currents are ANP4 for
$j_\sigma$ and VP1 for $j_x^{(1)}$. These currents have the form,
{\mathletters
\begin{equation}
\Delta j_\sigma[\mbox{ANP4}] \sim \int dx  
[ F_{202}^1(x) - \case{1}{4} D_{202}^1(x) ]
 \times j_0 (Q x/2) \, V_T^\pi(x)
\end{equation}
\begin{equation}
\Delta j_x^{(1)}[\mbox{VP1}] \sim \int dx 
D_{000}^1(x)
\times j_1 (Q x/2) \, x_\pi [ V_T^\pi (x) - V_S^\pi (x) ]
\end{equation}}
The above observation shows that the parts of the potential which are
most important to fit well are $V_T^\pi $ and $V_T^\pi - V_S^\pi$ in
the ranges where $(F_{202}^1 - \case{1}{4} D_{202}^1) j_0 (Q x/2)$ and
$D_{000}^1 \, j_1 (Q x/2) x_\pi$ are appreciable. The above
considerations led to the following practical procedure. We minimized
the function $f^{pq}(\Lambda_\rho, \Lambda_a)$ where, 
{\mathletters
\begin{equation}
f^{pq} = (I_T-I_T^0)^2 + (I_{TS} - I_{TS}^0)^2	
\end{equation}
and,
\begin{equation}
I_T = \int_{0}^{\infty} dx \, x^p \, V_T^{\rm OBEP}(x) 
\end{equation}
\begin{equation}
I_T^0 = \int_{0}^{\infty} dx \, x^p \, V_T^{\rm AV14}(x) 
\end{equation}
\begin{equation}
I_{TS} = \int_{0}^{\infty} dx \, x^q \, (V_{T}^{\rm OBEP}(x) 
- V_{S}^{\rm OBEP}(x)) 
\end{equation}
\begin{equation}
I_{TS}^0 = \int_{0}^{\infty} dx \, x^q \, (V_{T}^{\rm AV14}(x) 
- V_{S}^{\rm AV14}(x)) .
\end{equation}}

The choice $(p,q)=(3,3)$ makes the integrand in $I_T$ peak at $x=1.6
\mbox{ fm}$ and the integrand in $I_{TS}$ peak at $1.9 \mbox{ fm}$ and
gives the best matching of shape to the nuclear densities for ANP4 and
VP1.  Examples of fit values for $\Lambda_\rho,\Lambda_a$ are given in
Table \ref{OBEPfit}. We see that $\Lambda_\rho$ is smaller when
$\kappa_V$ is larger as one would expect. The value of $\Lambda_a$ is
poorly constrained and apparently may vary between $m_a$ and
$\Lambda_\rho$.

To summarize, $\Lambda_\rho$ and $\Lambda_a$ were chosen to match the
isospin dependent spin-spin and tensor potentials used to construct
the wavefunction. They are functions of $\Lambda_\pi,f_{\pi
NN},g_{\rho NN}$ and $\kappa_{V}$ and functionals of $V_{T}^{\rm
AV14}$ and $V_{S}^{\rm AV14}$. This somewhat artificial procedure
would be unnecessary if the wavefunctions were constructed using OBEP
potentials found from the same lagrangian as the exchange currents
were derived from. In that case $\Lambda_\pi,\Lambda_\rho$ and
$\Lambda_a$ would be fixed by the deuteron properties and NN
scattering data.  An alternative improvement which could be made is to
choose a different type of form factor, the idea being to match the
shape of $V_T^{AV14}$ and $V_S^{AV14}$ very closely.

There are four estimates of the $\pi N\Delta$ coupling $f_{\pi N
\Delta}$. The simplest constituent quark model yields $f_{\pi
N\Delta}= 6\sqrt{2}/5 f_{\pi NN}$ and hence $f_{\pi N\Delta}^2/4\pi
=0.23$. Dispersion theory yields $f_{\pi N\Delta}^2/4\pi =0.29$
\cite{TF}. The $\Delta$ width of 120 MeV implies $f_{\pi
N\Delta}^2/4\pi =0.35$ \cite{SM}. The highest estimate is $f_{\pi
N\Delta}^2/4\pi =0.36$ which is the value implied by relation $f_{\pi
N\Delta} = 3/\surd{2} f_{\pi NN}$ coming from the Skyrme-Soliton model
with $1/N_c$ corrections
\cite{Holz}. We note that this model also yields $f_{\pi
NN}^2/4\pi=0.080$ and $g_A=-1.28$ which are in good agreement with
experiment.

We take the range $f_{\pi N\Delta}^2/4\pi =\mbox{0.23 - 0.36}$
reflecting the various models: there is a large uncertainty in the
value of this parameter.

Our calculation does not treat the effect of the $\Delta$-isobar
explicitly i.e.\ there are no $\Delta$NN components in the
wavefunction. We were therefore unable to take into account direct
coupling to $\Delta$-isobars present in the nuclei or the indirect
effect of the $\Delta$-isobars on the coupling of the nucleons. These
processes have been considered for the axial current at zero momentum
transfer by Adam {\em et al.} \cite{AHHST}. The direct coupling
contributes +0.030 to the Gamow-Teller matrix element, M(GT), and the
indirect effect --0.021 \cite{HH}. These two effects compensate each
other although their sum, +0.009, is not insignificant.

A further deficiency of the calculation is the static approximation
for the propagator of a $\Delta$N pair. Hajduk {\em et al.} have shown
that this approximation is not valid and leads to an overestimate of
the $\Delta$-excitation current by a factor of 1.9 for both M(GT) and
$\mu_V$, \cite[Table 1]{HSY}. The reason for the overestimate is that
$M_\Delta-M$ is a poor estimate for the energy of the $NN\Delta$
configuration minus the energy of the NNN configuration. This is due
to the fact that $\Delta$-excitation occurs only in total orbital
angular momentum 2 channels: for these channels the kinetic energy is
large which contradicts the zero value given to it in the static
approximation.

To improve the static approximation we replaced $M_\Delta - M$ by
$M_\Delta - M + \langle T \rangle$ where $\langle T \rangle$ is some
average excess kinetic energy.  The value $\langle T \rangle = 110
\mbox{ MeV}$ was used, chosen so that the contribution of static
approximation $\Delta$-excitation to M(GT) found in Ref.\ \cite{HSY},
0.055, is converted to the exact value of 0.031 plus the contribution
from the $\Delta NN$ components, 0.009, i.e.
\begin{equation}
\frac{M_\Delta - M}{M_\Delta - M+\langle T \rangle} \times 0.055 = 0.040
\end{equation}
This procedure corrects the contribution of the $\Delta$-isobar
at zero momentum transfer but the $Q$ dependence of our $\Delta$
contribution is not correct. According to Fig.\ 3 of \cite{HSY}, we
overestimate the $\Delta$-excitation current for $Q<4 \mbox{ fm}^{-1}$
although the error is small at $Q=0.52 \mbox{ fm}^{-1}$ which is the
momentum transfer for nuclear muon capture by \mbox{$^{3}$He}.

Finally we list the values used for other constants entering the
calculation.  We have used the three-momentum transfer $Q=103.22
\mbox{ MeV}$, the energy transfer given by lepton kinematics $Q_0=
2.44 \mbox{ MeV}$, $m_\pi=138.03 \mbox{ MeV}$, $M_N = 939 \mbox{
MeV}$, $M_\Delta=1232 \mbox{ MeV}$, $m_\rho = 770 \mbox{ MeV}$, $m_\mu
= 105.66 \mbox{ MeV}$, $f_\pi= 92 \mbox{ MeV}$, $g_A(0)=-1.257\pm
0.003$, $g_V = 0.974\pm 0.001$, $g_M = 3.576\pm0.001$, $g_A=-1.236 \pm
0.003$ (the values for $g_V, g_M$ and $g_A$ are at $q^2 = -0.954
m_\mu^2$).

Besides the uncertainty in the parameters, we need to take into
account the fact that there are many possible realistic NN
potentials. Wavefunctions derived from these potentials will yield
different matrix elements according to the relative strengths of the
different parts of the potential e.g.\ tensor interaction. To take
this into account properly requires calculating the observables using
wavefunctions derived from all the different potentials. We were not
able to do this but we did estimate the `model uncertainty' by varying
the size of the dominant matrix element densities by a constant factor
i.e.\ independent of $x$. The size of this factor was determined as
described below.

The one-body currents are dominated by $[{\bf 1}]^0$ and
$[{\bbox \sigma}]^{0,1}$. The variation in the value of
$[{\bf 1}]^0$ can be neglected since all models will agree at $Q=0$
\footnote{The value is one minus a correction of the order of few
$\times 10^{-4}$ due to isospin symmetry breaking} and will have the
same deviation from that value at low $Q$ provided the isovector
radius is reasonable. This condition will be satisfied for
wavefunctions derived from realistic potentials which possess the
correct binding energy because of scaling \cite{FGCP}.

The above assertion was tested by calculating $[{\bbox 1}]^0$ at
$Q=103$ MeV with the AV14 and AV14+3BF 8 channel wavefunctions which
have different binding energies. The bessel functions entering the
matrix element can be expanded at small $Q$ as,
\begin{equation}
j_0(2 Q y/3) = 1 - Q^2 y^2 /54 + \ldots
\end{equation}
and so the change, $\Delta$, in $[{\bbox 1}]^0$ as $Q$ changes from
zero to a small finite value is $Q^2 <y^2> /54$ which scales as
$1/E_B$. If we use the mean \mbox{$^{3}$He}/\mbox{$^{3}$H} binding
energy and the value $[{\bbox 1}]^0=0.851$ for the AV14+3BF
wavefunctions then scaling implies,
\begin{equation}
\Delta (AV14) = 0.149 \times \frac{(8.34+7.67)}{(7.66 + 7.01)} = 0.163.
\end{equation}
The above scaling argument predicts $[{\bbox 1}]^0=0.837$ for the AV14
wavefunctions which is close to the calculated value of 0.839: this
result supports the argument that the model dependence in the value of
$[{\bbox 1}]^0$ is small.

For $[{\bbox \sigma}]^{0,1}$, however, differences do exist at
$Q=0$. The impulse approximation contributions $\mu_V^{\rm IA}$ and
\mbox{M(GT)$^{\rm IA}$} are both proportional to $[{\bbox
\sigma}]^{0,1}$ which allowed us to gauge the variation in 
$[{\bbox \sigma}]^{0,1}$ due to the use of various wavefunctions.  We
used values for $\mu_V^{\rm IA}$ reported in \cite{LAmag} for the Reid
soft core (RSC), RSC+TM, RSC+Brazilian(BR), AV14, AV14+TM and AV14+BR
potentials. Schiavilla {\em et al.} \cite{SPR} report values for
$\mu_V^{\rm IA}$ for the AV14+Urbana VII and Urbana+Urbana VII
potentials. The value of $[{\bbox \sigma}]^{0,1}$ for the Paris
potential was taken from the calculation of M(GT) reported in
\cite{AHHST}. Calculations of M(GT) have also been performed for the
 Paris, super-soft core, AV14 and RSC potentials as reported in
Ref.~\cite{Saito}.  The range for $[{\bbox \sigma}]^{0,1}$ from these
sources is \mbox{(--0.913) - (--0.932)} and so in our calculation we
varied $[{\bbox \sigma}]^{0,1}$ by a factor ranging from 0.99 to
1.01. We did not include the large values of 0.937 and 0.943 reported
in Refs.~\cite{AHHST,Saito} for Bonn-type potentials in this analysis
because of their very different nature. Bonn-type potentials will
yield a peculiar balance between one-body and two-body currents:
basically less two-body and more one-body because of their weak tensor
force.

For the exchange currents the dominant matrix elements are
$D_{000}^1$, $F_{000}^1$, $D_{202}^1$ and $F_{202}^1$ which enter into
the $\Delta$-excitation current and the pion pair (with PS coupling)
or contact (with PV coupling) term. By studying the 34 channel entries
of Tables IV and II of Ref.~\cite{LAmag} we arrived at a common
variation factor of 0.917 - 1.083 for $D_{202}^1$ and $F_{202}^1$ and
0.941 - 1.059 for $D_{000}^1$ and $F_{000}^1$.

These variations due to `model dependence' made significant
contributions to the total uncertainty quoted as did the variations in
$f_{\pi N\Delta}$ and $\Lambda_\pi$. The uncertainties in the
observables were found by a Monte-Carlo analysis where the parameters
were chosen randomly from their ranges with a flat probability
distribution\footnote{In the case of experimental uncertainties the
range was taken as $\pm \sqrt{3} \sigma$}.  It was checked that the
probability distribution for the observables was close to a normal
distribution with the same mean and variance: one expects this because
of the central limit theorem.

The final results are shown in Table \ref{res1} and the uncertainties
listed are one standard deviation. Also shown are results for the 8
channel wavefunction: there is little change in going from 8 to 22
channels for the wavefunctions used here although we did notice
significant fractional changes in the value of the smaller matrix
elements, in particular those dominated by total orbital angular
momentum 1 states. We also include results found from the AV14
wavefunctions. The large difference in the rate is due to scaling
rather than being a direct consequence of the three-body force. If the
one-body currents are taken from the 22 channel AV14+3BF wavefunction
and different wavefunctions used for the two-body currents, then the
results shown in table \ref{res4} are obtained. These results are very
similar to each other showing that the exchange currents are less
sensitive to scaling than the one-body currents and that both one-body
and two-body currents are insensitive to the increase in the number of
channels in the CRC wavefunction.

In table \ref{res2} we list the contributions of each current to the
current amplitudes and effective form factors. The largest exchange
current corrections are from the axial delta-excitation currents
(ANP4-6) and the vector $\pi$-pair (contact) term (VP1). The agreement
between the microscopic calculation and the EPM is very good except
for $j_x^{(1)}$ and $j_x^{(2)}$ which differ by 7\% and 11\%
respectively. Table \ref{res3} shows the separate contributions of
local and non-local currents.

The dependencies of the observables on the nucleon pseudoscalar
coupling $g_P$ are shown in Fig.~\ref{figgp} and the sensitivies are
listed in Table~\ref{res5}. The dependence on $g_P$ is similar to that
on the trion pseudoscalar coupling $F_P$ found in
Ref.~\cite{JCHF}. The rate is less sensitive to $g_P$ than $A_v$ which
in turn is less sensitive than either $A_t$ or $A_\Delta$. The curves
shown in Fig.~\ref{figgp} are well reproduced by parameterizing the
effective couplings as follows and using Eqs.~(3),(11),(12) and (13)
of Ref.~\cite{JCHF}.  
{\mathletters
\begin{eqnarray}
G_V &=& 0.835 \\
G_P &=& 0.231 + 0.370  r  \\
G_A &=& 1.300
\end{eqnarray}}
where $r$ is the ratio of $g_P$ to its PCAC value.

Our result for $\mu_V$ is $-2.52 \pm 0.03$ which agrees with the
experimental value of --2.55.  Our result for M(GT) is $0.977 \pm
0.013$ which also agrees with the experimental value of $0.961 \pm 0.003$.

\section{Conclusions}
\label{concl}

The calculation presented here used nucleonic and mesonic degrees of
freedom to describe the charge changing weak nuclear current of the
trion system at low momentum transfer. The two-nucleon component of
the current is given by the $\pi$-MEC obtained from the hard-pion
lagrangian of the $N \Delta \pi \rho A_1$ system. The nuclear system
is described by wavefunctions derived by the coupled rearrangement
channel method from the AV14 NN potential with Tuscon-Melbourne
three-body force.

We first checked our numerics by calculating the trion isovector
magnetic moment and the Gamow-Teller matrix element.  The results of
Table~\ref{3Nmag} and Table~\ref{GT} agree well with the results
of Refs.~\cite{LAmag} and ~\cite{AHHST}, respectively.

Our analysis of the observables for reaction ~(\ref{reac}) shows that
the main contribution comes from the space-like component of the
current.

The potential $\pi$-MEC, connected with the nuclear OPEP via the
nuclear continuity equation, is relatively well defined because the
parameters of the OPEP are well known: every realistic potential
should respect them. This statement is weakened by the fact that the
AV14 potential is not of the OBEP type and the needed value of the
cutoff $\Lambda_{\pi}$ can be extracted only approximately
(Table~\ref{param}).  The axial part of the potential MEC (entries 2-7
of Table~\ref{res2}) contributes significantly to $G_P$, while its
contribution to $G_A$ is only a minor one, because of a destructive
interference of the individual contributions. The vector part of the
potential MEC (lines 14-15 of Table~\ref{res2}) contribute both to
$G_A$ and $G_P$, with the prevailing contribution from the pair term
VP1.

There are two sets of non-potential currents. One set is present
only in the weak axial MEC and arises due to the interaction of the
weak axial current with the $\pi \rho A_1$ system (currents
ANP1-3). The contribution of these currents to the observables is only
a minor one (lines 8-10 of Table~\ref{res2}).  The other set of
non-potential currents is formed by the $\Delta$-excitation currents
ANP4-6 and VNP1-2.  The axial currents ANP4-6 contribute to $G_A$
considerably (lines 11-13 of Table~\ref{res2}).  This set of currents
is much more model dependent and it is mainly responsible for the
uncertainty of the calculation. It follows from the analysis of
section III that the contribution from these currents is slightly
overestimated due to the static limit in which these current are
considered here.

One of our main goals is the analysis of the possibility of extraction
of $g_P$. We have found that the rate is rather insensitive to it but
the spin observables offer the opportunity of measuring $g_P$
precisely (see Fig.~\ref{figgp}). However, these kind of experiments are also
more difficult to perform.

We presented the final results for the observables in
Table~\ref{res1}. It can be seen that the microscopic calculations and
the EPM predictions agree very well. In particular, our result for the
transition rate for reaction (1.1) is, \begin{equation} \Gamma_0 =
1502 \pm 32\,s^{-1}\,. \label{Ga0th}
\end{equation}
Our estimate of the uncertainty in the calculation yields an error of
$\approx$ 2 \% in the value of the transition rate. The largest part
of the uncertainty in $\Gamma_0$ comes from poor knowledge of $f_{\pi
N \Delta}$ .  The value of $\Gamma_0$ Eq.\ (\ref{Ga0th}) is in a good
agreement with the preliminary results of the new precise
measurement~\cite{exp},
\begin{equation}
     \Gamma^{\rm exp}_0 = 1494 \pm 4 \,s^{-1}\,. \label{Ga0exp}
\end{equation}
Using this value we can compare (\ref{Ga0th}) and (\ref{Ga0exp}) and
conclude that
\begin{enumerate}
\item the structure of the space-like component of the weak axial
$\pi$-MEC is well described at low momentum transfer within the
framework of the phenomenological hard-pion method,
\item the value of the induced pseudoscalar constant $g_P$ is,
\end{enumerate}
\begin{equation}
\frac{g_P}{g_P^{\rm PCAC}} = 1.05 \pm 0.19
\end{equation}
and so the PCAC value of $g_P$ is in rough agreement with the data as
is the heavy baryon chiral perturbation theory value,
$g_P^{CPT}/g_P^{\rm PCAC} = 1.01 \pm 0.02$ \cite{Bern}. By $g_P^{\rm PCAC}$ we
mean the value found from,
\begin{equation}
g_P^{\rm PCAC}(q^2) = \frac{2m_\mu M}{m_\pi^2-q^2} g_A(q^2)
\end{equation}
which yields $g_P^{\rm PCAC}(-0.954 m_\mu^2)=8.12$. Let us note that
the best measurement of ordinary muon capture by the proton
\cite{Bard1} yields and uncertainty in $g_P$ of 42\% and combining
various measurements reduces this to 22\% \cite{Bard2}.

The 0.19 uncertainty in $g_P$ results almost entirely from the 2\%
theoretical uncertainty in the calculation of the rate. If the spin
analyzing power $A_v$ was measured and used to infer the value of
$g_P/g_P^{\rm PCAC}$ then the lower limit on the uncertainty set by
theory is 0.02.


\acknowledgments
One of the authors (JGC) would like to acknowledge the support of the
Institute for Theoretical Physics, Utrecht, the Netherlands where some
of this work was performed. We are grateful to M.Kamimura who provided
the coefficients for the nuclear wavefunctions. We would like to thank
J.L.Friar and E.L.Tomusiak for providing us with additional
information regarding the trion isovector magnetic moment calculation
of Ref.~\cite{LAmag} and J.Adam Jr.\ for useful discussions.

\appendix
\section{Momentum space currents}
\label{appc}

In this appendix we list the currents used in the calculation. They
are written in their momentum space representation. Our conventions
are that $g_A < 0$ and that the total current is the sum of the vector
current $j_V$ and the axial current $j_A$, not the vector current
minus the axial current. The overall sign of the currents are
consistent with the dominant one-body axial current,
\begin{equation}
j_A^a(1) = +g_A \vec{\sigma} \frac{\tau^a}{2}. \end{equation}

The one-body currents are listed below and are consistent with the
scheme used in~\cite{JCHF}. Here, the initial and final momenta of
nucleon $i$ are written $\vec{p}$ and $\vec{p}\,'$ respectively and
the Pauli spin matrices for nucleon $i$ are written $\vec{\sigma}$.
\begin{eqnarray}
\vec{j}_A^a &=&
\frac{\tau^a}{2} \Bigl\{
g_A \Bigl[ \vec{\sigma}
\Bigl( 1 - \frac{(\vec{p}\,'+\vec{p}\,)^2}{8M^2} \Bigr)
+\frac{1}{4M^2}\Bigl(i \vec{p} \times \vec{p}\,'
+ \vec{p}\,' (\vec{p} \cdot \vec{\sigma})
+ \vec{p} (\vec{p}\,' \cdot \vec{\sigma})
\Bigr) \Bigr] \nonumber \\
&& + g_P \frac{\vec Q}{m_\mu}
\Bigl[ \frac{\vec{\sigma} \cdot \vec{p}}{2M}
\Bigl( 1 - \frac{3\vec{p}\,^2}{8M^2} - \frac{\vec{p}\,'^2}{8M^2} \Bigr)
-\frac{\vec{\sigma} \cdot \vec{p}\,'}{2M}
\Bigl( 1 - \frac{\vec{p}\,^2}{8M^2} - \frac{3\vec{p}\,'^2}{8M^2} \Bigr)
\Bigr]\Bigr\} \end{eqnarray}
\begin{eqnarray}
\rho_A^a &=&
\frac{\tau^a}{2} \Bigl\{
g_A \frac{\vec{\sigma} \cdot (\vec{p}\,'+\vec{p})}{2M} \nonumber \\
&& + g_P \frac{Q_0}{m_\mu}
\Bigl[ \frac{\vec{\sigma} \cdot \vec{p}}{2M}
\Bigl( 1 - \frac{3\vec{p}\,^2}{8M^2} - \frac{\vec{p}\,'^2}{8M^2} \Bigr)
-\frac{\vec{\sigma} \cdot \vec{p}\,'}{2M}
\Bigl( 1 - \frac{\vec{p}\,^2}{8M^2} - \frac{3\vec{p}\,'^2}{8M^2} \Bigr)
\Bigr\}  \end{eqnarray}
\begin{eqnarray}
\vec{j}_V^a &=&
\frac{\tau^a}{2} \Bigl\{
g_V \Bigl[
\frac{\vec{p}+\vec{p}\,'}{2M}
+ \frac{i \vec{\sigma} \times (\vec{p}\,'-\vec{p}\,)}{2M} \Bigr]
+ g_M \Bigl[ \frac{i\vec{\sigma} \times \vec{Q}}{2M}
-\frac{Q_0}{2M} \Bigl(\frac{\vec{Q}}{2M} +
\frac{i \vec{\sigma} \times (\vec{p}\,'+\vec{p}\,)}{2M}
\Bigr)\Bigr] \Bigr\} \end{eqnarray}
\begin{eqnarray}
\rho_V^a &=&
\frac{\tau^a}{2} \Bigl\{
g_V \Bigl[ 1 - \frac{(\vec{p}\,'-\vec{p}\,)^2}{8M^2} +
\frac{i \vec{\sigma} \cdot (\vec{p}\,' \times \vec{p}\,)}{4M^2} \Bigr]
+ g_M \Bigl[ -\frac{(\vec{p}\,'-\vec{p}\,)^2}{4M^2} +
\frac{i \vec{\sigma} \cdot (\vec{p}\,' \times \vec{p}\,)}{2M^2}
\Bigr] \Bigr\}
\end{eqnarray}

Now follow the two-body currents which have been labeled AP$i$,
ANP$i$, VP$i$ or VNP$i$ for the sake of reference. A(V) stands for
axial-vector (vector) and P(NP) stands for potential (non-potential)
current. The currents are written in terms of non-local momenta
$\vec{P}_2,\vec{P}_3$ and local momenta $\vec{Q}_2,\vec{Q}_3$ defined
by,
\begin{eqnarray}
\vec{P}_2 &=& \vec{p}\,'_{\!2} + \vec{p}_2 \\
\vec{Q}_2 &=& \vec{p}\,'_{\!2} - \vec{p}_2 \\
\vec{P}_3 &=& \vec{p}\,'_{\!3} + \vec{p}_3 \\
\vec{Q}_3 &=& \vec{p}\,'_{\!3} - \vec{p}_3 \end{eqnarray}
where $\vec{p}_i^{\,'},(\vec{p}_i)$ is the momentum of nucleon $i$ in
the final (initial) state. The currents appearing below are for
the pair of particles labeled (23) and the isospin component $a \in
\{x,y,z\}$ i.e.\ we have written here $j^a(23)$. The total current for
muon capture is then $j^{x-iy}(12) + j^{x-iy}(21) + j^{x-iy}(23) +
j^{x-iy}(32) + j^{x-iy}(31) + j^{x-iy}(13)$.  Given the current
$j(23)$ for particles 2 and 3, the matrix element of the other five
other currents $j(32), j(12) \ldots$ are equal to that of $j(23)$
which follows from the symmetry of the wavefunctions under interchange
of particle labels.

First we list the weak axial potential currents;\\
the sum of the $a_1$-pole pair term and the PCAC constraint term,
\widetext
\begin{eqnarray}\label{AP1AP7}
\vec{j}_A^{a,\mbox{\scriptsize bare}}[\mbox{AP1+AP7}] &=&
\left( \frac{g}{2M} \right)^2
\frac{g_A}{2M}
\Delta_F^\pi(Q_3) F_{\pi NN}^2(Q_3)
(\vec{\sigma}_3 \cdot \vec{Q}_3) \Bigl\{
[\vec{Q}+i\vec{\sigma}_2\times \vec{P}_2] \tau_3^a 
\nonumber \\ &&
+ [\vec{P}_2 + i \vec{\sigma}_2 \times \vec{Q}]
i (\tau_2 \times \tau_3)^a
\Bigr\} 
\end{eqnarray}
where the total current is related to the bare current for this
current and currents AP3,AP4 and APSPV by,\footnote{In the calculation,
the axial form factor $F_A(Q)$ was set to one so that the second term
in Eq.\ (\ref{AP1AP7t}) yielded nothing.  The correct value of $F_A(Q)$
is one minus 0.019.}
\begin{equation} \label{AP1AP7t}
\vec{j}_A^a =
F_A(Q) F_{\pi NN}(Q) \vec{j}_A^{a,\mbox{\scriptsize bare}} 
+F_{\pi NN}(Q) (1 - F_A(Q)) \frac{\vec{Q}}{Q^2-Q_0^2}
(\vec{Q} \cdot \vec{j}_A^{a,\mbox{\scriptsize bare}}),
\end{equation}
the $\pi$-pole pair term plus $\pi$-pole contact term,
\widetext
\begin{eqnarray} \label{AP2AP6}
\vec{j}_A^{a} [\mbox{AP2+AP6}]&=&
 \left( \frac{g}{2M} \right)^2  \frac{g_A}{4M}
\left( \frac{2 m_\pi^2}{Q^2 - Q_0^2 + m_\pi^2} \right) F_{\pi NN}(Q)
\Delta_F^\pi(Q_3) F_{\pi NN}^2(Q_3)
\vec{Q} \frac{(\vec{\sigma}_3 \cdot \vec{Q}_3)}{m_\pi^2}
\nonumber \\&& 
\hspace{-2ex}
\times \Bigl\{
[\vec{Q}^2 + \vec{Q}_3^2
+ i\vec{\sigma}_2 \cdot (\vec{P}_2 \times \vec{Q}_2) ]
 \tau_3^a 
\nonumber \\ && 
\hspace{-2ex}
 + [ (\vec{Q} \cdot \vec{P}_2 )
 + 4 (\vec{Q}_3 \cdot \vec{P}_2)
 - 3 (\vec{Q}_3 \cdot \vec{P}_3)
 + 3 i \vec{\sigma}_2 \cdot (\vec{Q} \times \vec{Q}_3)]
i (\tau_2 \times \tau_3)^a \Bigr\}
\end{eqnarray}
the $\pi$-retardation term,
\begin{eqnarray} \label{AP3}
\vec{j}_A^{a,\mbox{\scriptsize bare}} \mbox{[AP3]}&=&
 -\left( \frac{g}{2M} \right)^2 \frac{g_A}{4M}
[\Delta_F^\pi(Q_3)]^2 F_{\pi NN}^2(Q_3) 
 (\vec{\sigma}_3 \cdot \vec{Q}_3) \nonumber \\
&& \times \Bigl\{
(1-\nu) (\vec{Q} \cdot \vec{Q}_3) \vec{
[Q}_3 \tau_3^a + i \vec{\sigma}_2 \times \vec{Q}_3
i (\tau_2 \times \tau_3)^a ]
\nonumber \\
&&+ [(1-\nu) \vec{P}_2 \cdot \vec{Q}_3
+ (1+\nu) \vec{P}_3 \cdot \vec{Q}_3 \vec{]
[Q}_3 i \vec{(\tau}_2 \times \vec{\tau}_3)^a
+ i \sigma_2 \times Q_3 \tau_3^a]
\Bigr\} 
\end{eqnarray}
the $\pi$-form factor term,
\begin{eqnarray} \label{AP4}
\vec{j}_A^{a,\mbox{\scriptsize bare}} [\mbox{AP4}] &=&
\left( \frac{g}{2M} \right)^2 \frac{g_A}{4M}
\Delta_F^\pi(Q_3) \frac{d}{dQ_3^2}F_{\pi NN}^2(Q_3) \nonumber \\
&& \times ( \vec{\sigma}_3 \cdot \vec{Q}_3) \Bigl\{
(1-\eta) (\vec{Q} \cdot \vec{Q}_3 )
[\vec{Q}_3 \tau_3^a + i \vec{\sigma}_2 \times \vec{Q}_3
i (\tau_2 \times \tau_3)^a ]
\nonumber\\
&&+ [(1-\eta) \vec{P}_2 \cdot \vec{Q}_3
+ (1+\eta) \vec{P}_3 \cdot \vec{Q}_3 \vec{]
[Q}_3 i \vec{(\tau}_2 \times \vec{\tau}_3)^a
+ i \sigma_2 \times Q_3 \tau_3^a]
\Bigr\},
\end{eqnarray}
the current which measures the difference between exchange currents
derived using pseudoscalar and pseudovector $\pi$N coupling,
\begin{eqnarray} \label{APSPV}
\vec{j}_A^{a,\mbox{\scriptsize bare}} [\mbox{APSPV}] &=&
- ( 1 - \lambda )
\left( \frac{g}{2M} \right)^2 \frac{g_A}{4M}
\Delta_F^\pi(Q_3) F_{\pi NN}^2(Q_3) \nonumber \\
&& \times \Bigl\{
\Bigl[[\vec{P}_2 + i \vec{\sigma}_2 \times \vec{Q}]
(\vec{\sigma}_3 \cdot \vec{Q}_3)
- \vec{Q}_3 (\vec{\sigma}_3 \cdot \vec{P}_3 ) \Bigr]
i (\tau_2 \times \tau_3)^a \nonumber\\
&& +\Bigl[[\vec{Q} + i \vec{\sigma}_2 \times \vec{P}_2 ]
( \vec{\sigma}_3 \cdot \vec{Q}_3 )
- i \vec{\sigma}_2 \times \vec{Q}_3 ( \vec{\sigma}_3 \cdot \vec{P}_3 )
\Bigr]
\tau_3^a \Bigr\},
\end{eqnarray}
the $a_1$-pole contact terms, normal plus anomalous,
\begin{eqnarray}\label{AP5}
\vec{j}_A^{a,\mbox{\scriptsize bare}} [\mbox{AP5}] &=&
-\left( \frac{g}{2M} \right)^2 \frac{1}{2g_A}
\Delta_F^\pi(Q_3) F_{\pi NN}(Q_3)  F_{\rho NN}(Q_2)
\nonumber \\ &&
 \times  \frac{( \vec{\sigma}_3 \cdot \vec{Q}_3)}{2M}
\Bigl\{
\vec{P}_2
+ (1+\kappa_V) i \vec{\sigma}_2 \times \vec{Q}_2 \Bigr\}
i (\tau_2 \times \tau_3)^a \end{eqnarray}
\begin{eqnarray}
\rho_A^{a,\mbox{\scriptsize bare}} [\mbox{AP5}]&=&
-\left( \frac{g}{2M} \right)^2 \frac{1}{2g_A}
\Delta_F^\pi(Q_3) F_{\pi NN}(Q_3)  F_{\rho NN}(Q_2)
\nonumber \\
&& \times ( \vec{\sigma}_3 \cdot \vec{Q}_3)
\Bigl\{ 1 + \frac{(1+2\kappa_V)}{8M^2} [
-\vec{Q}_2^2 + i \vec{\sigma}_2 \cdot ( \vec{Q}_2 \times \vec{P}_2 )]
\Bigr\} i (\tau_2 \times \tau_3)^a,
\end{eqnarray}
where the relationship between the total and bare current in this case
is\footnote{Again, the $F_A(Q) = 1$ approximation was made so that
the second term in Eq.\ (\ref{AP5t}) yielded nothing.}
\begin{equation} \label{AP5t}
\vec{j}_{A\mu}^a =
F_A(Q) F_{\pi NN}(Q) \vec{j}_{A\mu}^{a,\mbox{\scriptsize bare}} +
F_{\pi NN}(Q) (1 - F_A(Q)) \frac{Q_\mu}{Q^2-Q_0^2}
(\vec{Q} \cdot \vec{j}_A^{a,\mbox{\scriptsize bare}}
- Q_0 \rho_A^{a,\mbox{\scriptsize bare}}).
\end{equation}

Here follow the weak axial non--potential currents. They are;\\
the $a_1 \rho\pi$ term,
\begin{eqnarray} \label{ANP1}
\vec{j}_A^{a,\mbox{\scriptsize bare}} [\mbox{ANP1}]&=&
- \left( \frac{g}{2M} \right)^2
\frac{1}{8M g_A}
\Delta_F^\rho (Q_2) \Delta_F^\pi (Q_3)
F_{\rho NN}(Q_2) F_{\pi NN}(Q_3)
( \vec{\sigma}_3 \cdot \vec{Q}_3 ) \nonumber \\
&&\times 
\Bigl\{
\Bigl[
\vec{P}_2
+ ( 1 + \kappa_V ) i \vec{\sigma}_2 \times \vec{Q}_2
\Bigr]
\Bigl[ \vec{Q}_2 \cdot ( \vec{Q}_3 - \vec{Q}_2 )
+ \frac{ \vec{P}_2 \cdot \vec{Q}_2}{M^2}
( \vec{P}_2 \cdot \vec{Q}_2 - \vec{P}_3 \cdot \vec{Q}_3 )\Bigr]
 \nonumber \\
&&  + \vec{Q}_2 \Bigl[ \vec{Q}_3 \cdot ( \vec{P}_3 - \vec{P}_2 )
- (1+\kappa_V) i \vec{\sigma}_2 \cdot ( \vec{Q}_2 \times \vec{Q}_3 )
\nonumber \\
&& + \frac{(1+2\kappa_V)}{8M^2} (\vec{P}_2 \cdot \vec{Q}_2)
[-\vec{Q}_2^2 + i \vec{\sigma}_2 \cdot ( \vec{Q}_2 \times \vec{P}_2 )]
 \Bigr]
\Bigr\}i (\tau_2 \times \tau_3)^a
\end{eqnarray}
\begin{eqnarray}
\rho_A^{a,\mbox{\scriptsize bare}} [\mbox{ANP1}]&=&
- \left( \frac{g}{2M} \right)^2
\frac{1}{4 g_A}
\Delta_F^\rho (Q_2) \Delta_F^\pi (Q_3)
F_{\rho NN}(Q_2) F_{\pi NN}(Q_3)
( \vec{\sigma}_3 \cdot \vec{Q}_3 ) \nonumber\\
&& \times \Bigl\{ \Bigl[
\vec{Q}_2 \cdot ( \vec{Q}_3 - \vec{Q}_2 )
+ \frac{( \vec{Q}_2 \cdot \vec{P}_2 )^2}{4 M^2}
\Bigr]
\Bigl[
1 + \frac{(1+2\kappa_V)}{8M^2}
[-\vec{Q}_2^2 + i \vec{\sigma}_2 \cdot ( \vec{Q}_2 \times \vec{P}_2 )]
\Bigr]
\nonumber \\
&& - \frac{( \vec{Q}_2 \cdot \vec{P}_2 )}{4 M^2}
[ ( \vec{Q}_2 \cdot \vec{P}_2 )
+ ( 1 + \kappa_V ) i \vec{\sigma}_2 \cdot \vec{(Q}_2 \times \vec{Q}_3) ]
\Bigr\} i (\tau_2 \times \tau_3)^a
\end{eqnarray}
where the relationship between total and bare current is given by
Eq.\ (\ref{AP5t}),\\
the $\rho\pi$ term,
\begin{eqnarray} \label{ANP2}
\vec{j}_A^a [\mbox{ANP2}]&=&
- \left( \frac{g}{2M} \right)^2
\frac{m_\rho^2}{4M g_A}
F_{\pi NN}(Q)
\Delta_F^\rho (Q_2) \Delta_F^\pi (Q_3)
F_{\rho NN}(Q_2) F_{\pi NN}(Q_3)
i (\tau_2 \times \tau_3)^a  
\nonumber \\
&&\times (\vec{\sigma}_3 \cdot \vec{Q}_3) \Bigl\{
 \vec{P}_2 + (1+\kappa_V) i \vec{\sigma}_2 \times \vec{Q}_2
\Bigr\}\end{eqnarray}
\begin{eqnarray}
\rho_A^a [\mbox{ANP2}] &=&
- \left( \frac{g}{2M} \right)^2
\frac{m_\rho^2}{2 g_A}
F_{\pi NN}(Q)
\Delta_F^\rho (Q_2) \Delta_F^\pi (Q_3)
F_{\rho NN}(Q_2) F_{\pi NN}(Q_3)
i (\tau_2 \times \tau_3)^a  \nonumber \\
&& \times (\vec{\sigma}_3 \cdot \vec{Q}_3) \Bigl\{
1 + \frac{(1+2\kappa_V)}{8M^2}
[-\vec{Q}_2^2 + i \vec{\sigma}_2 \cdot ( \vec{Q}_2 \times \vec{P}_2 )]
\Bigr\},
\end{eqnarray}
the $\rho\pi\pi$ term,
\begin{eqnarray} \label{ANP3}
\vec{j}_A^a [\mbox{ANP3}]&=&
-\vec{Q}
\left( \frac{g}{2M} \right)^2
\frac{m_\rho^2}{2M g_A^2}
\Delta_F^\pi(Q)
\Delta_F^\rho (Q_2) \Delta_F^\pi (Q_3)
F_{\rho NN}(Q_2) F_{\pi NN}(Q_3) \nonumber \\
&& \times (\vec{\sigma}_3 \cdot \vec{Q}_3) \Bigl\{
 \vec{Q}_3 \cdot ( \vec{P}_3 - \vec{P}_2 ) - ( 1 + \kappa_V )
i \vec{\sigma}_2 \cdot ( \vec{Q}_2 \times \vec{Q}_3 )
\nonumber \\
&&  + (\vec{Q}_3 \cdot \vec{P}_3) \frac{(1+2\kappa_V)}{8M^2}
[-\vec{Q}_2^2 + i \vec{\sigma}_2 \cdot ( \vec{Q}_2 \times \vec{P}_2 )]
\Bigr\}i (\tau_2 \times \tau_3)^a. \end{eqnarray}
delta-excitation, $\pi$ propagator,
\begin{eqnarray} \label{ANP4}
\vec{j}_A^a [\mbox{ANP4}]&=&
 g_A \frac{f_{\pi N\Delta}^2}{m_\pi^2}
\frac{4}{9( M_\Delta -M )}
\Delta_F^\pi(Q_3) F_{\pi NN}^2(Q_3) 
\nonumber \\ &&
 \times ( \vec{\sigma}_3 \cdot \vec{Q}_3)
\Bigl\{ \vec{Q}_3 \tau_3^a
+ \case{1}{4} i ( \vec{\sigma}_2 \times \vec{Q}_3 )
i (\tau_2 \times \tau_3)^a \Bigr\}, 
\end{eqnarray}
delta-excitation, $\rho$ propagator,
\begin{eqnarray} \label{ANP5}
\vec{j}_A^a [\mbox{ANP5}]&=&
-\left(\frac{f_{\pi N\Delta}}{m_\pi}\right) \frac{(1+\kappa_V)}
{2M^2} \frac{4G_1 g_\rho^2 f_\pi}{9(M_\Delta - M)}
\Delta_F^\rho(Q_3) F_{\rho NN}^2(Q_3) \nonumber \\
&&\times \Bigl\{ \vec{Q}_3 \times ( \vec{\sigma}_3 \times \vec{Q}_3 ) \tau_3^a
+ \case{1}{4}
i \vec{\sigma}_2 \times ( \vec{Q}_3 \times ( \vec{\sigma}_3
\times \vec{Q}_3 )) i (\tau_2 \times \tau_3 )^a \Bigr\},
\end{eqnarray}
delta-excitation, $a_1$ propagator,
\begin{equation}
\vec{j}_A^a [\mbox{ANP6}]=
\left( \frac{f_{\pi N\Delta}}{f_{\pi NN}} \right)^2 g_A^3
g_\rho^2 \frac{4}{9(M_\Delta - M)}
\Delta_F^a(Q_3) F_{aNN}^2(Q_3)  
 \Bigl\{ \bar{\sigma}_3 \tau_3^a +
\case{1}{4} i \vec{\sigma}_2 \times \bar{\sigma}_3
i (\tau_2 \times \tau_3)^a
\Bigr\}
\end{equation}
where,
\begin{equation} \bar{\sigma}_3 = \vec{\sigma}_3
+ \frac{\vec{Q}_3 ( \vec{\sigma}_3 \cdot \vec{Q}_3 )}{M_{a_1^2}}. 
\end{equation}

Finally we list the weak vector currents. They are;\\
the $\pi$-pair term, (with PS coupling) or $\pi$-contact term
(with PV coupling),
\begin{eqnarray} \label{VP1}
\vec{j}_V^a [\mbox{VP1}]&=&
- \left( \frac{f_{\pi NN}}{m_\pi} \right)^2
\Delta_F^\pi(Q_3) F_{\pi NN}^2(Q_3)
 \vec{\sigma}_2 (\vec{\sigma}_3 \cdot \vec{Q}_3 )
i (\tau_2 \times \tau_3)^a,
\end{eqnarray}
the pion current,
\begin{eqnarray} \label{VP2}
\vec{j}_V^a [\mbox{VP2}]&=&
+ \left( \frac{f_{\pi NN}}{m_\pi} \right)^2
\vec{Q}_2( \vec{\sigma}_2 \cdot \vec{Q}_2 )
( \vec{\sigma}_3 \cdot \vec{Q}_3 ) i (\tau_2 \times \tau_3)^a
\nonumber \\ &&
\times \frac{1}{Q^2_2-Q^2_3}
\left[\Delta_F^\pi(Q_3) F^2_{\pi NN}(Q_3) 
- \Delta_F^\pi(Q_2) F^2_{\pi NN}(Q_2)\right],
\end{eqnarray}
delta-excitation, $\pi$ propagator,
\begin{eqnarray} \label{VNP1}
\vec{j}_V^a [\mbox{VNP1}]&=&
-\frac{2G_1}{M} \frac{4f_{\pi N\Delta} f_{\pi NN}}{9m_\pi^2(M_\Delta -M)}
\Delta_F^\pi(Q_3) F_{\pi NN}^2(Q_3)
\nonumber \\ &&
\times ( \vec{\sigma}_3 \cdot \vec{Q}_3 )
 i \vec{Q} \times
\Bigl\{ \vec{Q}_3 \tau_3^a
+ \case{1}{4} i ( \vec{\sigma}_2 \times \vec{Q}_3 )
i (\tau_2 \times \tau_3)^a \Bigr\},
\end{eqnarray}
delta-excitation, $\rho$ propagator,
\begin{eqnarray} \label{VNP2}
\vec{j}_V^a [\mbox{VNP2}]&=&
- \frac{(1+\kappa_V)}{2M} \left( \frac{G_1}{M} \right)^2
\frac{4g_\rho^2}{9(M_\Delta -M)}
\Delta_F^\rho(Q_3) F_{\rho NN}^2(Q_3) \nonumber\\
&&\times 
i \vec{Q} \times
\Bigl\{ \vec{Q}_3 \times ( \vec{\sigma}_3 \times \vec{Q}_3 ) \tau_3^a
+ \case{1}{4}
i \vec{\sigma}_2 \times ( \vec{Q}_3 \times ( \vec{\sigma}_3
\times \vec{Q}_3 )) i (\tau_2 \times \tau_3 )^a \Bigr\}.
\end{eqnarray}
\narrowtext

\section{Transformation into configuration representation}
\label{apptrans}

 Purely local currents of the form $f(Q_2,Q_3) g(Q_3)$ were
transformed into configuration representation according to,
{\mathletters
\begin{eqnarray}
\tilde{\jmath}_{23}(\vec{x},\vec{y}\,) &=& 
\exp [ i \vec{Q}\cdot(\case{\vec{x}}{2}-\case{\vec{y}}{3})]
\int \frac{d^3q}{(2\pi)^3}
\exp[-i\vec{q}\cdot\vec{x}\,] 
\times f(\vec{Q}-\vec{q},\vec{q}\,) 
\, g(\vec{q}\,)\\
&=& 
\exp [ i \vec{Q}\cdot(\case{\vec{x}}{2}-\case{\vec{y}}{3}\,)]
f(\vec{Q}-i\grave{\nabla}_x,i\grave{\nabla}_x) \tilde{g}(x)
\end{eqnarray}}
where,
{\mathletters
\begin{equation}
\vec{x} = \vec{r}_2-\vec{r}_3 \end{equation}
\begin{equation}  
\vec{y} = \vec{r}-\frac{1}{2}(\vec{r}_2+\vec{r}_3)\end{equation}
\begin{equation}
\tilde{g}(x) = \int \frac{d^3q}{(2\pi)^3} \exp[-i\vec{q}\cdot\vec{x}\,]
g(\vec{q}\,)
\end{equation}}
and $\grave{\nabla}_x$ is a derivative acting only on the function
immediately to its right, in this case $\tilde{g}(x)$.

This process yields an operator in configuration space with a plane
wave factor, a spin-space part made from $\vec{\sigma}_2,\vec{\sigma}_3$
and $\hat{x}$, an isospin part and a potential function which is a
function of $x=\mid \vec{x} \mid$. As an example consider the vector
current $\pi$ pair term.
\begin{eqnarray}
\vec{j}_V^a(\vec{Q}_2,\vec{Q}_3) &=& -\left( \frac{f_{\pi NN}}{m_\pi}
\right)^2 i(\tau_2 \times \tau_3)^a 
\vec{\sigma}_2 (\vec{\sigma}_3
\cdot \vec{Q}_3) \Delta_F^\pi(Q_3) F_{\pi NN}^2(Q_3) \end{eqnarray}
\begin{eqnarray} \Rightarrow
\vec{\tilde{\jmath}}_V^{\,a}(\vec{x},\vec{y}\,) &=& +f_{\pi NN}^2
\exp [ i \vec{Q}\cdot(\case{\vec{x}}{2}-\case{\vec{y}}{3})]
i(\tau_2 \times \tau_3)^a 
i \vec{\sigma}_2 (\vec{\sigma}_3 \cdot \hat{x}) f_1(x)
\end{eqnarray}

The potential function in this case is $f_1(x)$, (see Fig.\ \ref{f1}),
the isospin operator is $i(\tau_2 \times \tau_3)^a$ and the plane wave
factor is $\exp [ i \vec{Q}\cdot(\vec{x}/2-\vec{y}/3)]$.

Non-local currents $(\alpha \vec{P}_2 + \beta \vec{P}_3)
f(\vec{Q}_2,\vec{Q}_3) g(\vec{Q}_3)$ were transformed to,
\begin{eqnarray}
\tilde{\jmath}_{23}(\vec{x},\vec{y}\,) &=& 
\exp [ i \vec{Q}\cdot(\case{\vec{x}}{2}-\case{\vec{y}}{3})]
\{ f(\vec{Q}-i\grave{\nabla}_x,i\grave{\nabla}_x) \tilde{g}(x) 
\nonumber \\
&&\hspace{-3ex} 
\times [\alpha(-2i\nabla_x+i\nabla_y) + \beta ( 2i\nabla_x+i\nabla_y)]
+[\alpha(\vec{Q}-i\grave{\nabla}_x)+\beta i\grave{\nabla}_x] 
f(\vec{Q}-i\grave{\nabla}_x,i\grave{\nabla}_x) \tilde{g}(x)\}
\nonumber \\
&&\end{eqnarray} 
An example of this type of term is part of the AP1+AP7 current.
\begin{eqnarray}
\vec{j}_A^a(\vec{Q}_2,\vec{Q}_3,\vec{P}_2,\vec{P}_3) 
&=& -\left( \frac{f_{\pi NN}}{m_\pi}\right)^2 
\frac{g_A}{2M}
i(\tau_2 \times \tau_3)^a 
\times \vec{P}_2 (\vec{\sigma_3}\cdot
\vec{Q}_3) \Delta_F^\pi(Q_3) F_{\pi NN}^2(Q_3) 
\nonumber \\
\end{eqnarray}
\begin{eqnarray}
\Rightarrow \tilde{\jmath}_A^a(\vec{x},\vec{y}) &=& -f_{\pi NN}^2
g_A \frac{m_\pi}{2M} 
\exp [ i \vec{Q}\cdot(\case{\vec{x}}{2}-\case{\vec{y}}{3})]
i(\tau_2 \times \tau_3)^a  
\times \{i(\vec{\sigma_3}\cdot\hat{x}) f_1(x) 
[-2i\vec{\nabla}_x+i\vec{\nabla}_y] 
\nonumber \\ && 
 + \frac{Q}{m_\pi} \hat{Q} i(\vec{\sigma_3}\cdot\hat{x}) f_1(x) 
-\hat{x} (\vec{\sigma_3}\cdot\hat{x}) f_2(x) + \vec{\sigma_3} f_3(x)\}
\end{eqnarray}

The operator $\vec{\nabla}$ was written as the sum of a part with a
derivative $\hat{x} \partial / \partial x$ and an angular part
$\hat{\nabla}_x/x$ where,
\begin{equation}
\hat{\nabla}_x = \hat{\theta}_x \frac{\partial} {\partial \theta_x} +
\frac{\hat{\phi}_x}{\sin \theta_x} \frac{\partial} {\partial \phi_x}
\end{equation}

\section{Leading contributions from each current}
\label{applead}

Here we write the contributions which were included from each current.
The argument of the two-body matrix elements are potential functions
$f_i$, (defined in appendix \ref{apppot}), multiplied by either
$j_0(Qx/2)$ which is suppressed for succinctness or $j_i (i\neq0)$
which is then written, e.g. $_xB_{000}^1[f_1]$ means,
\begin{eqnarray}
\langle \, \mbox{$^{3}$H} \mid \mid 
12 /\surd 3 [ 4\pi {\cal Y}_{00}^0(\hat{x}\hat{y}) 
\otimes S_0(\vec{\sigma}_2,\vec{\sigma}_3)]_1 
i ( \tau_2 \times \tau_3 )^-
 f_1(x) j_0 ( Q x / 2) j_0( Q y /3) 
\frac{\partial}{\partial x_\pi}
\mid \mid \mbox{$^{3}$He} \, \rangle
\end{eqnarray}
and $F_{112}^1[j_2 f_2]$ means,
\begin{eqnarray}
\langle \, \mbox{$^{3}$H} \mid \mid 
12 / \surd 3 [ 4\pi {\cal Y}_{11}^2(\hat{x}\hat{y}) 
\otimes \vec{\sigma}_3]_1 \tau_3^- 
 f_2(x) j_2 ( Q x / 2) j_1 ( Q y /3) 
\mid \mid \mbox{$^{3}$He} \, \rangle.
\end{eqnarray}
The one-body matrix elements $[{\bf 1}]^0$, 
$[{\bbox \sigma}]^\pm$ and $[{\bbox \sigma}]^{l,1}$ 
are defined by Eqs.\ (47-50) of \cite{JCHF}.\\
\underline{One-body current}
\begin{equation}
\Delta \rho_0 = \Bigl [ g_V \Bigl ( 1 - \frac{Q^2}{8M^2} \Bigl  ) 
- g_M \frac{Q^2}{4M^2} \Bigr ] [{\bf 1}]^0
 \end{equation}
\begin{equation}
\Delta \rho_1 = -\frac{Q}{2M} \Bigl [ -\frac{g_A}{3} +
g_P \frac{q^0}{m_\mu} \Bigl ( 1 - \frac{Q^2}{24M^2} \Bigr ) \Bigr ]
[{\bf \sigma}]^+  \end{equation}
\begin{equation}
\Delta j_Q = \frac{Q}{2M} \Bigl [ \frac{g_V}{3}  - g_M \frac{q^0}{2M} 
\Bigr ] [{\bf 1}]^0 \end{equation}
\begin{equation}
\Delta j_x^{(1)} = -\frac{Q}{2M} \Bigl [ 
g_V + g_M \Bigl ( 1 - \frac{q^0}{6M} \Bigr ) \Bigr ] [{\bbox \sigma}]^- 
 \end{equation}
\begin{eqnarray}
\Delta j_\sigma &=& g_A \Bigl ( 1 + \frac{Q^2}{24M^2} \Bigr )
[{\bbox \sigma}]^{0,1} 
-\frac{1}{3} 
\frac{Q}{2M} \Bigl [ g_A \frac{Q}{3M} + g_P \frac{Q}{m_\mu} \Bigl (
1 - \frac{Q^2}{24M^2} \Bigr ) \Bigr ] [{\bbox \sigma}]^+  \end{eqnarray}
\begin{eqnarray}
\Delta j_x^{(2)} &=& 
\frac{3}{\surd 2} g_A \Bigl ( 1 + \frac{Q^2}{24M^2} \Bigr )
[{\bbox \sigma}]^{2,1}
-\frac{Q}{2M} \Bigl [ g_A \frac{Q}{3M} + g_P \frac{Q}{m_\mu} \Bigl (
1 - \frac{Q^2}{24M^2} \Bigr ) \Bigr ] [{\bbox \sigma}]^+
\end{eqnarray}
\underline{AP1 + AP7}
\begin{eqnarray}
\Delta j_\sigma &=& - g_A f_{\pi NN}^2 F_{\pi NN}(Q)
   \frac{m_\pi}{M} \times \Bigl\{
     \case{1}{3}           _xB_{000}^1[f_1]
    -\case{\surd 2}{3}     _xB_{202}^1[f_1]
    +\case{\surd 2}{3}     _xH_{000}^1[f_1]        
    +\case{1}{3}     _xH_{202}^1[f_1]   
\nonumber\\
&&  +\case{1}{\surd 3}     _xI_{202}^1[f_1]
    -                      _xB_{101}^{011}[f_1]
    -\surd \case{2}{9}     _xG_{101}^{111}[f_1]
    -\surd \case{2}{3}     _xH_{101}^{111}[f_1]   
    -\surd \case{10}{9}    _xI_{101}^{111}[f_1]\Bigr\}
\end{eqnarray}
\widetext
\underline{AP2+AP6}
\begin{eqnarray}
\Delta j_x^{(2)} &=& 3\Delta j_\sigma = -g_A f_{\pi NN}^2  F_{\pi NN}(Q)
\frac{2m_\pi^2}{(Q^2-Q_0^2+m_\pi^2)} 
\frac{Q}{2M} \frac{Q}{m_\pi} 
\nonumber \\&&
\times \Bigl\{ 
   -\case{1}{\surd 2}    D_{000}^1[f_2-3f_3]
   -\case{1}{2}          D_{202}^1[f_2]
 +\case{1}{3}         _xB_{000}^{1}[f_1]
   -\case{\surd 2}{3}   _xB_{202}^{1}[f_1] 
   +\case{\surd 2}{3}   _xH_{000}^{1}[f_1]
\nonumber \\&&
 +\case{1}{3}         _xH_{202}^{1}[f_1]
   +\case{1}{\surd 3}   _xI_{202}^{1}[f_1]
   -                    _xB_{101}^{011}[f_1]
  -\surd \case{2}{9}  _xG_{101}^{111}[f_1]
   -\surd \case{2}{3}   _xH_{101}^{111}[f_1]  
\nonumber \\ && 
   -\surd \case{10}{9}  _xI_{101}^{111}[f_1]\Bigr\}
\end{eqnarray}
\underline{AP3}
\begin{eqnarray}
\Delta j_\sigma &=& - \nu g_A f_{\pi NN}^2 F_{\pi NN}(Q) 
\frac{m_\pi}{M} \times \Bigl\{
    \case{1}{3}         _xB_{000}^1[5f_{20}-f_{19}]
   -\case{\surd 2}{3}   _xB_{202}^1[2f_{20}-f_{19}] 
\nonumber \\&&
   +\case{\surd 2}{3}   _xH_{000}^1[5f_{20}-f_{19}]
   +\case{1}{3}         _xH_{202}^1[2f_{20}-f_{19}]
   +\case{1}{\surd 3}   _xI_{202}^1[2f_{20}-f_{19}] 
   -\case{4}{3}         _xB_{101}^{011}[f_{20}]
\nonumber \\&&
   +\case{1}{\surd 3}   _xB_{101}^{111}[f_{20}]
   -\case{\surd 5}{3}   _xB_{101}^{211}[f_{20}] 
   +\case{\surd 2}{3}   _xH_{101}^{011}[f_{20}]
   -\surd \case{3}{2}   _xH_{101}^{111}[f_{20}]
\nonumber \\&&
   -\surd \case{5}{18}  _xH_{101}^{211}[f_{20}] 
   -\surd \case{5}{2}   _xI_{101}^{111}[f_{20}]
   +\surd \case{5}{6}   _x I_{101}^{211}[f_{20}]\Bigr\}
\end{eqnarray}
\underline{AP4}
the same as for AP3 with 
\begin{eqnarray}
                       \nu &\rightarrow& -2\eta \nonumber \\
                      f_{19} &\rightarrow& f_{21} \nonumber \\
                      f_{20} &\rightarrow& f_{22} 
\end{eqnarray}
\underline{APSPV}
\begin{eqnarray}
\Delta j_\sigma &=& 
-(1-\lambda) g_A f_{\pi NN}^2 F_{\pi NN}(Q)   \frac{m_\pi}{M} 
\times \Bigl\{
   \case{-1}{3}           _xB_{000}^1[f_1]
   +\case{\surd 2}{3}     _xB_{202}^1[f_1]
   -\case{\surd 2}{3}     _xH_{000}^1[f_1] 
\nonumber \\&&
   -\case{1}{3}           _xH_{202}^1[f_1]
   -\case{1}{\surd 3}     _xI_{202}^1[f_1]
   +\case{2}{3}           _xB_{101}^{011}[f_1]
   -\case{1}{2\surd 3}    _xB_{101}^{111}[f_1] 
\nonumber \\&&
   +\surd \case{5}{36}    _xB_{101}^{211}[f_1]
   -\surd \case{2}{36}    _xH_{101}^{011}[f_1]
   +\surd \case{3}{8}     _xH_{101}^{111}[f_1]
\nonumber \\&&
   +\surd \case{5}{72}    _xH_{101}^{211}[f_1] 
   +\surd \case{5}{8}     _xI_{101}^{111}[f_1]
   -\surd \case{5}{24}    _xI_{101}^{211}[f_1]\Bigr\}
\end{eqnarray}
\underline{AP5}
\begin{eqnarray}
 \Delta j_\sigma &=& -\frac{f_{\pi NN}^2}{g_A} F_{\pi NN}(Q) 
\frac{ m_\pi}{2M} 
\times \Bigl\{ \frac{(1+\kappa_V)}{2}
\Bigl[    \surd 2     D_{000}^1[f_5/3-f_6]
   +\case{1}{3}       D_{202}^1[f_5]\Bigr ]
\nonumber \\ &&
   + 
\Bigl[
   -\case{1}{3}       _xB_{000}^1[f_4]
   +\case{\surd 2}{3} _xB_{202}^1[f_4]
   +                  _xB_{101}^{111}[f_4]
\Bigr ] \Bigr\}
\end{eqnarray}
\begin{equation}
\Delta \rho_1 = G_3 \Bigl\{
 \case{\surd 2}{3} _xD_{000}^1[f_4]
+\case{1}{3}       _xD_{202}^1[f_4]
-\surd \case{2}{3} _xD_{101}^{111}[f_4] \Bigr \}
\end{equation}
where,
\begin{equation}
G_3 = (-) \frac{f_{\pi NN}^2}{g_A} \frac{m_\pi Q}{8M^2} 
(1+2\kappa_V) F_{\pi NN}(Q).
\end{equation}
\underline{ANP1}
\begin{eqnarray}
\Delta j_\sigma &=& -\frac{f_{\pi NN}^2}{g_A} 
\frac{m_\pi^2}{2 M m_\rho}
\times \Bigl\{
\frac{(1+\kappa_V)}{2} \Bigl[
  \surd 2 D_{000}^1[f_{15}/3-f_{16} - \case{7}{6}(\case{Q}{m}_\pi)^2 
(f_{12}/3-f_{13})]    
\nonumber \\ &&
+\case{1}{3}D_{202}^1[f_{15} 
- \case{7}{6}(\case{Q}{m}_\pi)^2 f_{12}]
\Bigr] 
   +\case{1}{3}   _xB_{000}^1[x_\pi f_{16}+5f_{18}-f_{17}
   -\case{1}{2}(\case{Q}{m}_\pi)^2 f_{11}]
\nonumber \\ &&
   -\case{\surd 2}{3}    _xB_{202}^1[x_\pi f_{16}+2f_{18}-f_{17}
   -\case{1}{2}(\case{Q}{m}_\pi)^2 f_{11}]
\nonumber \\&&
   -                   _xB_{101}^{011}[x_\pi f_{16}+\case{4}{3}f_{18}
   -\case{1}{2}(\case{Q}{m}_\pi)^2 f_{11}]
   +\case{1}{\surd 3}  _xB_{101}^{111}[f_{18}]
   -\case{\surd 5}{3}  _xB_{101}^{211}[f_{18}]\Bigr\}
\end{eqnarray}
\begin{eqnarray}
\Delta j_x^{(2)} &=& -\frac{f_{\pi NN}^2}{g_A}
(1+\kappa_V)
\frac{Q^2}{16 M m_\rho} 
\times \Bigl\{
   \surd 2        D_{000}^1[f_{12}/3-f_{13}]
   +\case{1}{3}   D_{202}^1[f_{12}] \Bigr\}
\end{eqnarray}
\begin{eqnarray}
\Delta \rho_1 &=& (-) G_3 \Bigl\{
 \case{\surd 2}{3} _xD_{000}^1[f_4]
+\case{1}{3}       _xD_{202}^1[f_4]
-\surd \case{2}{3} _xD_{101}^{111}[f_4] \Bigr\} 
\nonumber \\&&
+G_3 \frac{m_\rho}{m_\pi} \Bigl ( 1 + \frac{Q^2}{2m_\rho^2} \Bigr ) 
\Bigl \{
 \case{\surd 2}{3} _xD_{000}^1[f_{11}]
+\case{1}{3}       _xD_{202}^1[f_{11}]
-\surd \case{2}{3} _xD_{101}^{111}[f_{11}] \Bigr\} 
\nonumber \\&&
+G_3 \frac{m_\pi}{2m_\rho} \Bigl \{
+                   _xD_{202}^1[f_{18}]
\hspace{-1ex}-\case{1}{\surd 6} _xD_{101}^{011}[f_{18}] 
+\case{1}{\surd 10} _xD_{101}^{111}[f_{18}] 
-\case{1}{\surd 15} _xD_{101}^{211}[f_{18}] \Bigr\}
\nonumber \\&&
+\frac{1+\kappa_V}{1+2\kappa_V} G_3 \frac{m_\pi}{m_\rho}
\Bigl \{
                   _xD_{000}^1[5f_{18}-f_{17}]
+\case{1}{3}       _xD_{202}^1[2f_{18}-f_{17}]
+\case{\surd 2}{3} _xD_{101}^{011}[f_{18}] 
\nonumber \\&&
-\surd  \case{3}{2} _xD_{101}^{111}[f_{18}]
-\case{1}{3}\surd \case{5}{2} _xD_{101}^{211}[f_{18}] \Bigr\}
\nonumber \\&&
-\frac{1+\kappa_V}{1+2\kappa_V} G_3 \frac{m_\rho}{2 m_\pi} \Bigl \{
\case{\surd 2}{3}  _xD_{000}^1[3f_{13}-f_{12}-\case{m_\pi}{m_\rho}(3f_6-f_5)]
-\case{1}{3}       _xD_{202}^1[f_{12}-\case{m_\pi}{m_\rho}f_5]
 \Bigr \}
\end{eqnarray}
\underline{ANP2}
\begin{eqnarray}
\Delta j_\sigma &=& -\frac{f_{\pi NN}^2}{g_A} 
\frac{(1+\kappa_V) m_\rho}{4M} F_{\pi NN}(Q) \times \Bigl\{
   \surd 2  D_{000}^1[f_{12}/3-f_{13}]
   +\case{1}{3}     D_{202}^1[f_{12}] \Bigr\} 
\nonumber \\&&
-\frac{f_{\pi NN}^2}{g_A} \frac{m_\rho}{2M}
F_{\pi NN}(Q) \times \Bigl\{
   -\case{1}{3}  _xB_{000}^1[f_{11}]
   +\case{\surd 2}{3}  _xB_{202}^1[f_{11}]
   +    _xB_{101}^{011}[f_{11}]\Bigr\}
\end{eqnarray}
\begin{eqnarray}
\Delta \rho_1 &=& G_3 \frac{m_\rho}{m_\pi}\Bigl\{
 \case{\surd 2}{3} _xD_{000}^1[f_{11}]
+\case{1}{3}       _xD_{202}^1[f_{11}]
-\surd \case{2}{3} _xD_{101}^{111}[f_{11}]
\end{eqnarray}
\underline{ANP3}
\begin{eqnarray}
\Delta j_x^{(2)} &=& 3\Delta j_\sigma = 
-\left(\frac{f_{\pi NN}}{g_A}\right)^2 
\frac{(1+\kappa_V) m_\rho Q^2}{2M(m_\pi^2-Q_0^2+Q^2)} 
\times 
\Bigl\{   \surd 2  D_{000}^1[f_{12}/3-f_{13}]
   +1/3    D_{202}^1[f_{12}]\Bigr\}
\nonumber \\&&
-\left(\frac{f_{\pi NN}}{g_A}\right)^2 
\left(\frac{m_\pi}{M}\right)^2\
\frac{(1+2\kappa_V) m_\rho Q^2}{4M(m_\pi^2-Q_0^2+Q^2)}
\times \Bigl\{
   \case{1}{3}  _xB_{000}^1[f_{17}-5f_{18}] 
\nonumber \\&&
   +\case{\surd 2}{3}  _xB_{202}^1[2f_{18}-f_{17}]
   +\case{4}{3}  _xB_{101}^{011}[f_{18}]
   -\case{1}{\surd 3}  _xB_{101}^{111}[f_{18}]
   +\case{\surd 5}{3}    _xB_{101}^{211}[f_{18}]\Bigr\}
\end{eqnarray}
\underline{ANP4}
\begin{eqnarray}
 \Delta j_\sigma &=&
g_A f_{\pi N\Delta}^2 \frac{4m_\pi}{9(M_\Delta-M)} \times  \Bigl\{
     F_{000}^1[f_3-f_2/3]
   +\case{\surd 2}{3}   F_{202}^1[f_2] 
\nonumber \\&&
   +\case{\surd 2}{4}   D_{000}^1[f_3-f_2/3] 
   -\case{1}{12}        D_{202}^1[f_2]\Bigr\}
\end{eqnarray}
\underline{Higher order terms for ANP4}
\begin{eqnarray}
 \Delta j_x^{(2)} &=& 
g_A f_{\pi N\Delta}^2 \frac{4m_\pi}{9(M_\Delta-M)} 
\times \Bigl\{
              F_{000}^1[j_2 f_2]
   -\case{3}{\surd 2}   F_{202}^1[j_2 (2f_2/3-f_3)] 
   -\case{3}{\surd 2}   F_{022}^1[j_0 (f_2/3-f_3)]
\nonumber \\&&
   +\case{2\surd 3}{5}  F_{110}^1[j_1 f_2]
   +\surd \case{27}{5}  F_{110}^1[j_1 (17f_2/30 - f_3)] 
   +\case{1}{\surd 5}   F_{220}^1[j_0 f_2]
   +\case{3}{2\surd 5}  F_{221}^1[j_0 f_2]
\nonumber \\ &&
   +\surd\case{7}{20}   F_{222}^1[j_0 f_2] 
   -\case{1}{4\surd 2}  D_{000}^1[j_2 f_2]
   -\case{3}{4}         D_{202}^1[j_2 (f_2/6-f_3)]
   -\case{3}{4}         D_{022}^1[j_0 (f_2/3-f_3)] 
\nonumber \\ &&
   +\case{9}{40} \surd \case{3}{10}   E_{111}^1[j_1 f_2]
   +\case{21}{40}  \surd \case{1}{10}     E_{110}^1[j_1 f_2]\Bigr\}
\end{eqnarray}
\begin{eqnarray}
 \Delta j_\sigma &=&
g_A f_{\pi N\Delta}^2 \frac{4m_\pi}{9(M_\Delta-M)} \times \Bigl\{
   \surd 3               F_{110}^1[j_1 (f_2/3-f_3)]
   +\surd \case{4}{15}   F_{112}^1[j_1 f_2]
\nonumber \\ && 
   -\surd \case{1}{40}   E_{112}^1[j_1 f_2]\Bigr\}
\end{eqnarray}
\narrowtext
\underline{ANP5}
\begin{eqnarray}
 \Delta j_\sigma = f_{\pi N\Delta} \frac{2 G_1 g_\rho^2 (1+\kappa_V) 
f_\pi}{9(M_\Delta-M)} \left(\frac{m_\pi}{M}\right)^2 
&\times &
\Bigl\{
                F_{000}^1[f_7-f_8/3]
   +\case{\surd 2}{3}   F_{202}^1[f_8] 
\nonumber \\&&
   +\case{\surd 2}{4}   D_{000}^1[f_7-f_8/3]
   -\case{1}{12}        D_{202}^1[f_8]\Bigr\}
\end{eqnarray}
\underline{ANP6}
\begin{eqnarray}
 \Delta j_\sigma &=& \left( \frac{f_{\pi N\Delta}}{f_{\pi NN}} \right)^2
g_\rho^2 g_A^3
\frac{4 m_\pi^3}{9m_a^2(M_\Delta-M)} 
\times \Bigl\{      F_{000}^1[F_8+f_{24}-f_{23}/3]
   +\case{\surd 2}{3}   F_{202}^1[f_{23}] 
\nonumber \\ && 
   +\case{\surd 2}{4}   D_{000}^1[F_8+f_{24}-f_{23}/3]
   -\case{1}{12}        D_{202}^1[f_{23}]\Bigr\}
\end{eqnarray}
\underline{VP1}
\begin{eqnarray}
\Delta j_x^{(1)} = f_{\pi NN}^2 \times \Bigl\{
&&   \case{1}{\surd 2}   D_{000}^1[j_1 f_1]
   +\case{1}{2}   D_{202}^1[j_1 f_1]
\nonumber \\ && 
   +\case{1}{\surd 6}   C_{111}^1[j_0 f_1]
   -\surd \case{5}{24}   E_{111}^1[j_0 f_1] 
   -\surd \case{5}{8}     E_{110}^1[j_0 f_1]\Bigr\}
\end{eqnarray}
\underline{VP2}
\begin{equation}
\Delta j_x^{(1)} = f_{\pi NN}^2 \frac{Q}{m_\pi} \Bigl\{
   \case{1}{\surd 2}   D_{000}^1[e_3/3-e_2]
   +\case{1}{6}     D_{202}^1[e_3]\Bigr\}
\end{equation}
\underline{VNP1}
\begin{eqnarray}
\Delta j_x^{(1)} = -f_{\pi N\Delta} f_{\pi NN} 
\frac{8 G_1 m_\pi Q}{9M (M_\Delta-M)} \times \Bigl\{
&&      F_{000}^1[f_3-f_2/3]
   +\case{\surd 2}{3}   F_{202}^1[f_2] 
\nonumber \\ && 
   +\case{\surd 2}{4}   D_{000}^1[f_3-f_2/3]
   -\case{1}{12}    D_{202}^1[f_2]\Bigr\}
\end{eqnarray}
\underline{VNP2}
\begin{eqnarray}
 \Delta j_x^{(1)} = 
\frac{2 G_1^2 g_\rho^2 (1+\kappa_V)\, m_\pi}{9(M_\Delta-M)} 
\left(\frac{m_\pi}{M}\right)^2 \frac{Q}{M} 
&&\times \Bigl\{
      F_{000}^1[f_7-f_8/3]
   +\case{\surd 2}{3}   F_{202}^1[f_8] 
\nonumber \\ && 
   +\case{\surd 2}{4}   D_{000}^1[f_7-f_8/3]
   -\case{1}{12}     D_{202}^1[f_8]\Bigr\}
\end{eqnarray}

\section{Definition of potential functions}
\label{apppot}

Here we define the potential functions $f_i$. They are derivatives of
root potential functions $F_i$, e.g.\ $f_1 = -dF_1/dx_\pi, f_2 =
d^2F_1/dx_\pi^2  - 1/x_\pi dF_1/dx_\pi $. We first define the root potential
functions: 
{\mathletters
\begin{eqnarray}
4\pi F_1 &=& 
Y_0 (x_\pi) 
- R_\pi^{\case{1}{2}} Y_0(x_{\Lambda_\pi})
- \case{1}{2}  R_\pi^{-\case{1}{2}} (R_\pi -1 ) Y(x_{\Lambda_\pi}) \\
4\pi F_2 &=& 
\case{1}{2} Y (x_\pi) 
+ \case{1}{2}  R_\pi^{-\case{1}{2}} Y(x_{\Lambda_\pi})
+ \frac{2}{R_\pi-1} [R_\pi^{\case{1}{2}} Y_0(x_{\Lambda_\pi})-Y_0(x_\pi)]) \\
4\pi F_3 &=& 
  \frac{1}{R_\pi-1} [R_\pi^{\case{1}{2}} Y_0(x_{\Lambda_\pi}) -Y_0(x_\pi)]
+ \case{1}{2} R_\pi^{-\case{1}{2}} Y(x_{\Lambda_\pi}) 
+ \case{1}{8} R_\pi^{-\case{1}{2}} (R_\pi-1) 
   x^2_{\Lambda_\pi} Y_1(x_{\Lambda_\pi})\\
4\pi F_4 &=& b[\Lambda_\rho,m_\rho] 
\Bigl \{ 
\frac{1}{b[\Lambda_\rho,m_\pi]} Y_0(x_\pi)
- \frac{R_{\pi}^{\case{1}{2}} 
}{b[\Lambda_\rho,\Lambda_\pi]} Y_0(x_{\Lambda_\pi})
+\frac{b[\Lambda_\pi,m_\pi] R_\rho^{\case{1}{2}}}
{b[\Lambda_\rho,\Lambda_\pi] b[\Lambda_\rho,m_\pi]}
 Y_0(x_{\Lambda_\rho}) \Bigr \} \\
4\pi F_6 &=& \Bigl ( \frac{m_\rho}{m_\pi}\Bigr )^3
\Bigl \{
Y_0(x_\rho) 
- R_\rho^{\case{1}{2}} Y_0(x_{\Lambda_\rho}) 
- \case{1}{2} R_\rho^{-\case{1}{2}} (R_\rho-1) Y(x_{\Lambda_\rho})
\Bigr \} \\
4\pi F_7 &=& m_\rho m_\pi
\Bigl \{
\frac{b[\Lambda_\rho,m_\rho] R_\pi^{\case{1}{2}}}
{b[\Lambda_\rho,\Lambda_\pi] b[\Lambda_\pi,m_\rho]}
Y_0(x_{\Lambda_\pi})
+ \frac{b[\Lambda_\rho,m_\rho]}{b[\Lambda_\rho,m_\pi] b[m_\rho,m_\pi]}
Y_0(x_\pi) 
\nonumber \\ && 
- \frac{m_\rho b[\Lambda_\pi,m_\pi] R_\rho^{\case{1}{2}}}
{m_\pi b[\Lambda_\rho,\Lambda_\pi] b[\Lambda_\rho,m_\pi]}
Y_0(x_{\Lambda_\rho})
- \frac{m_\rho b[\Lambda_\pi,m_\pi]}
{m_\pi b[\Lambda_\pi,m_\rho] b[m_\rho,m_\pi]}
Y_0(x_\rho) \Bigr \}\\
4\pi F_8 &=& \Bigl ( \frac{m_{a_1}}{m_\pi}\Bigr )^3
\Bigl \{ Y_0 (x_{a_1}) 
- R_{a_1}^{\case{1}{2}} Y_0(x_{\Lambda_{a_1}})
- \case{1}{2}  R_{a_1}^{-\case{1}{2}} (R_{a_1} -1 ) Y(x_{\Lambda_{a_1}})
\Bigr \} 
\end{eqnarray}}
where $R_M = (\Lambda_M/m_M)^2$ and $Y(x) = \exp(-x)$, $Y_0(x) =
Y(x)/x$, $Y_1(x) = Y_0(x) (1+1/x)$, $Y_2(x)=Y_0(x) (1+3/x+3/x^2)$,
$Y_3(x)=Y_0(x)(1+6/x+15/x^2+15/x^3)$,
$b[\alpha,\beta]=\alpha^2-\beta^2$, $x_\pi = m_\pi x$,
$x_{\Lambda_\pi}= \Lambda_\pi x$, $x_\rho = m_\rho x$,
$x_{\Lambda_\rho} = \Lambda_\rho x$, $x_{a_1} = m_{a_1} x$ and
$x_{\Lambda_{a_1}} = \Lambda_{a_1} x$.

We now define the $f_i$ in terms of the root potential functions with
rules like, 
{\mathletters
\begin{equation}
f_8 = F_6[ Y_0 \rightarrow Y_2, Y 
\rightarrow y Y_1; 1/m_\rho^2].
\nonumber \end{equation}
The rule means that $f_8$ is equal to $F_6$ with $Y_0$ replaced by
$Y_2$ and $Y(\xi)$ replaced by $\xi Y_1(\xi)$ where $\xi$ can be any
argument. Further, each term $\exp(-a x)$ becomes $(a/m_\rho)^2
\exp(-a x)$ which is indicated by the $1/m_\rho^2$ after the
semi-colon.
\begin{eqnarray}
f_1 &=& F_1 [ Y_0 \rightarrow Y_1, Y \rightarrow Y ; 1/m_\pi] \\
f_2 &=& F_1[ Y_0 \rightarrow Y_2, Y \rightarrow y Y_1; 1/m_\pi^2] \\
f_3 &=& f_1 / x_\pi\\
\nonumber \\
f_4 &=& F_4[ Y_0 \rightarrow Y_1; 1/m_\pi] \\
f_5 &=& F_4[ Y_0 \rightarrow Y_2; 1/m_\pi^2] \\
f_6 &=& f_4 / x_\pi \\
\nonumber \\
f_7 &=& F_6[ Y_0 \rightarrow Y_0+Y_1/y, 
Y \rightarrow  Y-Y_0; 1/m_\rho^2] \\
f_8 &=& F_6[ Y_0 \rightarrow Y_2, Y 
\rightarrow y Y_1; 1/m_\rho^2] \\ 
\nonumber \\
f_{11} &=& F_7[ Y_0 \rightarrow Y_1; 1/m_\pi]\\
f_{12} &=& F_7[ Y_0 \rightarrow Y_2; 1/m_\pi^2]\\
f_{13} &=& f_{11}/x_\pi\\
f_{14} &=& F_7[ Y_0 \rightarrow Y_1+Y_2/y; 1/m_\pi^3]\\
f_{15} &=& F_7[ Y_0 \rightarrow Y_2; 1/m_\pi^4]\\
f_{16} &=& F_7[ Y_0 \rightarrow Y_1/y; 1/m_\pi^4]\\
f_{17} &=& F_7[ Y_0 \rightarrow Y_3; 1/m_\pi^3]\\
f_{18} &=& f_{12}/x_\pi \\
\nonumber \\
f_{19} &=& F_2[ Y_0 \rightarrow Y_3, 
Y \rightarrow y Y_2; 1/m_\pi^3] \\
f_{20} &=& F_2[ Y_0 \rightarrow Y_2/y, 
Y \rightarrow Y_1; 1/m_\pi^3] \\
\nonumber \\
f_{21} &=& F_3[ Y_0 \rightarrow Y_3, Y \rightarrow y Y_2,
              y^2 Y_1 \rightarrow y^2 Y_1; 1/m_\pi^3] \\
f_{22} &=& F_3[ Y_0 \rightarrow Y_2/y, Y \rightarrow Y_1,
              y^2 Y_1 \rightarrow  Y; 1/m_\pi^3] \\
\nonumber \\
f_{23} &=& F_8[ Y_0 \rightarrow Y_2, 
Y \rightarrow y Y_1; 1/m_{a_1}^2] \\
f_{24} &=& F_8[ Y_0 \rightarrow Y_1/y, 
Y \rightarrow Y_0; 1/m_{a_1}^2] \\
\nonumber \\
f_{25} &=& F_2 [ Y_0 \rightarrow Y_1, Y \rightarrow Y ; 1/m_\pi]\\
f_{26} &=& F_2[ Y_0 \rightarrow Y_2, Y \rightarrow y Y_1; 
1/m_\pi^2].
\end{eqnarray}}

The functions $e_i$ are given by,
\begin{equation}
8\pi e_2 = \frac{1}{x_\pi} [ e_{11}(m_\pi) - e_{11}(\Lambda_\pi) 
+ (R_\pi - 1) e_{12}(\Lambda_\pi) ] \end{equation}
\begin{equation}
8\pi e_3 =  e_{31}(m_\pi) - e_{31}(\Lambda_\pi) 
+ (R_\pi - 1) e_{32}(\Lambda_\pi) \end{equation}
where,
\begin{equation}
e_{11}(m) = \int_0^1 dt \, \exp(-c x) [ j_0 + \case{Q t}{c} j_1 ] \end{equation}
\begin{equation}
e_{12}(m) = -\frac{m_\pi^2}{2} \int_0^1 dt \, \exp(-c x) [ \case{x}{c} j_0
+\case{Q t}{c^3} (1+c x) j_1 ] \end{equation}
\begin{eqnarray}
e_{31}(m) &=& \frac{1}{m_\pi} \int_0^1 dt \, \exp(-c x) [ 
(\case{1}{x} + c - \case{Q^2 t^2}{c} ) j_0 
+ Q t ( \case{3}{c x} + 2 ) j_1 ] \end{eqnarray}
\begin{eqnarray}
e_{32}(m) &=& -\frac{m_\pi}{2} \int_0^1 dt \, \frac{\exp(-c x)}{c} [ 
(c x - \case{Q^2 t^2}{c^2} (1+c x) ) j_0
+ \case{Q t}{c} (2 c x+ 3 + \case{3}{c x}) j_1 ], \end{eqnarray} 
the argument of the bessel functions is $Q t x$ and the $m$ dependence
comes from $c$ which is given by
\begin{equation}
c(m) = [ Q^2 t (1-t) + m^2 ]^{\case{1}{2}}. \end{equation}

The functions $d_i$  are given by,
\begin{eqnarray}
d_2 = \case{1}{2} F_1 
d_3 = \case{1}{2} x_\pi f_1 
d_5 = \case{1}{2} x_\pi f_2. \end{eqnarray}

\section{Construction of OBEP}
\label{appOBEP}

The exchange of $\pi,\rho$ and $a_1$ mesons leads to an isospin
dependent NN potential with tensor and spin-spin components. The
non-relativistic momentum space potentials between two nucleons
labeled 1 and 2 were taken to be,
{\mathletters
\begin{equation}
V^\pi = - 3 c_\pi \, \tau_1 \cdot \tau_2 \, 
\frac{(\sigma_1 \cdot Q_2) (\sigma_2 \cdot Q_2)}{m_\pi^2}
\frac{\Delta_F^\pi(Q_2)}{m_\pi} F_{\pi NN}^2(Q_2) \end{equation}
\begin{eqnarray}
V^\rho &=& - 3 c_\rho \, \tau_1 \cdot \tau_2 \, 
\frac{(\sigma_1 \times Q_2) \cdot (\sigma_2 \times Q_2)}{m_\rho^2}
\times \frac{\Delta_F^\rho(Q_2)}{m_\rho} F_{\rho NN}^2(Q_2) 
\left ( \frac{m_\rho}{m_\pi} \right )^3
\end{eqnarray}
\begin{eqnarray}
V^{a_1} &=& 
- 3 c_{a_1} \, \tau_1 \cdot \tau_2 \, 
\left [ \sigma_1 \cdot \sigma_2+
\frac{(\sigma_1 \cdot Q_2) (\sigma_2 \cdot Q_2)}{m_{a_1}^2} \right ]
\times \frac{\Delta_F^{a_1}(Q_2)}{m_{a_1}} F_{a_1 NN}^2(Q_2) 
\left ( \frac{m_{a_1}}{m_\pi} \right )^3
\end{eqnarray}}
where,
{\mathletters
\begin{equation}
 c_\pi = c_{a_1} = m_\pi f_{\pi NN}^2/3
\end{equation}
\begin{equation}
 c_\rho = \frac{m_\pi}{3} g_{\rho NN}^2 (1 + \kappa_V)^2  
\left ( \frac{m_\pi}{2M} \right )^2.
\end{equation}}
The configuration representation of these potentials are, 
(with $M=\pi,\rho,a_1$),
\begin{equation}
V^M(x) = c_M [  V_T^M(x) S_{12} +  V_S^M(x) \sigma_1 \cdot \sigma_2]
\end{equation}
where,
\begin{equation}
S_{12} = 3 (\sigma_1 \cdot \hat{x}) (\sigma_2 \cdot \hat{x}) - 
\sigma_1 \cdot \sigma_2
\end{equation}
and,
{\mathletters
\begin{equation}
V_T^\pi(x) = F_\pi^{''} - F_\pi^{'}/x_\pi \end{equation}
\begin{equation}
V_T^\rho(x) = -(F_\rho^{''} - F_\rho^{'}/x_\rho) \end{equation}
\begin{equation}
V_T^{a_1}(x) = F_{a_1}^{''} - F_{a_1}^{'}/x_{a_1} \end{equation}
\begin{equation}
V_S^\pi(x) = F_\pi^{''} + 2 F_\pi^{'}/x_\pi  \end{equation}
\begin{equation}
V_S^\rho(x) = -2 (F_\rho^{''} + 2 F_\rho^{'}/x_\rho ) \end{equation}
\begin{equation}
V_S^{a_1}(x) = F_{a_1}^{''} + 2 F_{a_1}^{'}/x_{a_1} -3 F_{a_1} 
\end{equation}}
with,
\begin{equation}
F_M = \frac{1}{2\pi^2 x_M} \int_0^\infty \frac{q \sin (q x)}{q^2 + m_M^2} 
F_{MNN}^2(q) \left ( \frac{m_M}{m_\pi} \right )^3.
\end{equation}

The notation $F^{'}_M$ means $d/dx_M F_M$. With monopole form factors,
$F_MNN(q) = (\Lambda_M^2 - m_M^2)/(\Lambda_M^2+q^2)$ we found that,
\begin{eqnarray}
  4\pi \, V^M_T(x) &=& 
(-1)^{\delta(M,\rho)} \left ( \frac{m_M}{m_\pi} \right )^3
\times  [Y_2(m_M x) 
- R_M^{\case{3}{2}} Y_2(\Lambda_M x) 
\nonumber \\ && 
- R_M^{\case{1}{2}} (R_M-1) \,\Lambda_M x \, Y_1(\Lambda_M x)] \end{eqnarray}
\begin{eqnarray}
  4\pi \, V^\pi_S(x) &=& 
Y_0(m_\pi x) 
- R_\pi^{\case{1}{2}} Y_0(\Lambda_\pi x) 
- R_\pi^{-\case{1}{2}} (R_\pi-1) \, Y(\Lambda_\pi x) \end{eqnarray}
\begin{eqnarray}
  4\pi \, V^\rho_S(x) &=& 
(-2) \left ( \frac{m_\rho}{m_\pi} \right )^3
[ Y_0(m_\rho x) 
- R_\rho^{\case{1}{2}} Y_0(\Lambda_\rho x) 
- R_\rho^{\case{1}{2}} (R_\rho-1) \, Y(\Lambda_\rho x)] \end{eqnarray}
\begin{eqnarray}
  4\pi \, V^{a_1}_S(x) &=& 
(-2) \left ( \frac{m_{a_1}}{m_\pi} \right )^3 
[Y_0(m_{a_1} x) 
- R_{a_1}^{\case{1}{2}} Y_0(\Lambda_{a_1} x) 
\nonumber \\ && 
- \frac{R_{a_1}^{-\case{1}{2}}}{4} (R_{a_1}-1) (R_{a_1}-3) \, 
Y(\Lambda_{a_1} x)].
\end{eqnarray}

Our exchange potentials for $\pi$ and $\rho$ are the same as those
used by others \cite{Town,Bonn}. Our exchange potential for $a_1$
agrees with that of Ref.\ \cite{Town} but not with that of
Ref.\ \cite{TrSch}. The difference is that the $(\sigma_1 \cdot Q_2) (\sigma_2
\cdot Q_2)$ term was left out in \cite{TrSch}.  Here we include this
term because we found that it contributes non-negligibly: $V_T^a(x)$
is entirely due to the term and $V_S^a(x)$ would be $\approx 50\%$
smaller at $x=1/\Lambda_{a_1}$ and 50\% larger for $x \gg
1/\Lambda_{a_1}$ without the term.

\begin{figure}
\caption{Examples of nuclear densities. Points were also 
calculated at \protect$x\protect$=5.5,6.5 and 7.5 fm but
 are not shown.}
\label{densities}
\end{figure}

\begin{figure}
\caption{Isospin dependent spin-spin and tensor potentials from 
pion exchange and first and second derivatives
\protect$F_\pi^{'}\protect$ and \protect$F_\pi^{''}\protect$. The
solid line is for \protect$\Lambda_\pi = 1.2\protect$ GeV and the
dashed line for \protect$\Lambda_\pi = \infty\protect$. The constant
\protect$c_\pi\protect$ has been set to one for these graphs.}
\label{figOBEP}
\end{figure}

\begin{figure}
\caption{Variation of observables with the nucleon pseudoscalar 
coupling $g_P$.}
\label{figgp}
\end{figure}

\begin{figure}
\caption{Potential function for the 
\protect$\pi\protect$ pair term of the vector current.
The pion cutoff is \protect$\Lambda_\pi=1200\mbox{ Mev}\protect$}
\label{f1}
\end{figure}

\narrowtext
\begin{table}
\caption{Effective couplings in the EPM and IA and their difference. 
The uncertainty in $G_{\rm{P}}$ reflects experimental uncertainty only.}
\label{IAG}
\centering
\begin{tabular}{lcccccc}
Model & \multicolumn{2}{c}{\protect$G_{\rm{V}}$} & 
\multicolumn{2}{c}{\protect$G_{\rm{P}}$} 
& \multicolumn{2}{c}{$G_{\rm{\!A}}$} \\ \hline
EPM & \multicolumn{2}{c}{$0.85\pm0.01$} & 
\multicolumn{2}{c}{$0.603\pm 0.001$} & 
\multicolumn{2}{c}{$1.29\pm 0.01$} \\
IA  & 0.84 & --1\% & 0.523 & --13\% & 1.19 & 
--8\% \\
\end{tabular}                  
\end{table}

\begin{table}
\caption{Origin of contributions to effective couplings}
\label{IAj}
\centering
\begin{tabular}{lccc}
 & \protect$G_{\rm{V}}$ & 
\protect$G_{\rm{P}}$ & $G_{\rm{\!A}}$ \\ \hline
Dominant & $\rho_{\rm \,V}$ & 
$\vec{j}_{\rm A}$,$\vec{j}_{\rm V}$ & $\vec{j}_{\rm A}$ \\
Other      & $\vec{j}_{\rm V}$ &
 $\rho_{\rm A}$ & $\rho_{\rm A}$,$\vec{j}_{\rm V}$ \\
\end{tabular}                  
\end{table}

\begin{table} 
\caption{Labeling of reduced matrix elements. The
superscript in the isospin part means $( )^- = \case{1}{2} [( )^x -
i( )^y]$.}
\label{AI}
\begin{tabular}{ccc}
$Z$&$S_\Sigma$&${\cal I}$ \\
\hline
$A$&$\openone$&$12i (\tau_2 \times \tau_3)^-$\\
$B$&$\vec{\sigma}_3$&$12i (\tau_2 \times \tau_3)^-$\\
$C$&$[\sigma_2 \otimes \sigma_3]_0$&$12i (\tau_2 \times \tau_3)^-$\\
$D$&$[\sigma_2 \otimes \sigma_3]_1$&$12i (\tau_2 \times \tau_3)^-$\\
$E$&$[\sigma_2 \otimes \sigma_3]_2$&$12i (\tau_2 \times \tau_3)^-$\\
$F$&$\vec{\sigma}_3$&$12 \tau_3^-$\\
$G$&$[\sigma_2 \otimes \sigma_3]_0$&$12 \tau_3^-$\\
$H$&$[\sigma_2 \otimes \sigma_3]_1$&$12 \tau_3^-$\\
$I$&$[\sigma_2 \otimes \sigma_3]_2$&$12 \tau_3^-$\\
\end{tabular}
\end{table}

\widetext
\begin{table}
\caption{Sizes of matrix elements. The Bessel function in $y$ has been
expanded and only the first term kept except in the cases of 
$D_{000}^1,D_{202}^1,F_{000}^1$ and $F_{202}^1$. The dominant overlap 
describes the total orbital angular momentum of the wavefunction
components which contribute most to the matrix element.}
\label{MEint}
\begin{tabular}{cdddc}
Matrix element &\multicolumn{3}{c}{Integral}& Dominant overlap(s)\\
&\multicolumn{1}{c}{AV14+3BF[8]}&
\multicolumn{1}{c}{AV14[8]}&
\multicolumn{1}{c}{AV14+3BF[22]}&\\
\hline
$D_{000}^1$ &    9.62&  9.57&  9.63&SS\\
$G_{000}^0$ &    5.78&  5.96&  5.77&SS\\
$F_{000}^1$ &  --3.55&--3.53&--3.55&SS\\
$D_{202}^1$ &    2.03&   1.99&  2.05&SD\\ 
$F_{202}^1$ & --0.900&--0.867&--0.907&SD\\
$I_{202}^0$ & --0.066&--0.056&--0.082&SD,DD\\ 
$E_{110}^1$ &    0.20&   0.20&  0.20&SD\\
$I_{110}^0$ &  --0.15& --0.15&--0.15&SD\\
$F_{110}^1$ &    0.14&   0.14&  0.14&SD\\ 
$B_{111}^0$ &   0.057&  0.061&  0.054&DD\\
$G_{110}^0$ &   0.045&  0.053&  0.044&SS,DD\\ 
$A_{111}^1$ &   0.035&  0.037&  0.025&DD\\
$C_{111}^1$ & --0.024&--0.023&--0.031&DD\\ 
$F_{110}^1$ & --0.017&--0.018&--0.017&SS,DD\\
$E_{111}^1$ & --0.012&--0.014&--0.012&DD,SP\\
$H_{111}^0$ &  0.004& 0.003& 0.000 &SP\\
$D_{022}^1$ & --0.017&--0.017&--0.017&SD\\
$F_{022}^1$ &   0.005&  0.005&  0.005&SD\\
$I_{022}^0$ &--0.001&--0.001&--0.001&DD\\
$_xD_{000}^1$ &--7.00& --6.79&--7.02&SS\\
$_xB_{000}^1$ &--0.44& --0.45& --0.44&SS\\
$_xH_{000}^1$ &  0.31&   0.32&   0.31&SS\\
$_xD_{202}^1$ &--1.82& --1.77&--1.84&SD\\
$_xB_{202}^1$ &--0.49& --0.48&--0.49&SD\\
$_xH_{202}^1$ &  0.38&   0.38&  0.38&SD\\
$_xI_{202}^1$ & 0.023&  0.019&  0.023&SD\\ 
$_xD_{101}^{011}$ &    1.3&   1.2 &  1.3  &SD\\
$_xB_{101}^{011}$ &   0.94&  0.90 &  0.94 &SD\\
$_xH_{101}^{011}$ & --0.66&--0.62 &--0.66 &SD\\
$_xD_{101}^{111}$ &  --1.1& --1.1 &--1.1  &SD\\
$_xB_{101}^{111}$ & --0.79& --0.77&--0.79 &SD\\
$_xH_{101}^{111}$ &   0.55&   0.54&  0.55 &SD\\
$_xD_{101}^{211}$ &   0.29&   0.28&  0.29 &SD\\ 
$_xB_{101}^{211}$ &   0.19&   0.20&  0.19 &SD\\ 
$_xH_{101}^{211}$ & --0.14& --0.14&--0.13 &SD\\ 
$_xG_{101}^{111}$ &  0.033&  0.036&  0.032 &DD\\
$_xI_{101}^{111}$ &--0.021&--0.022&--0.21 &SD,DD\\
$_xI_{101}^{211}$ &--0.023&--0.024&--0.025 &DD\\
\end{tabular}
\end{table}

\narrowtext
\begin{table} 
\caption{Trion isovector magnetic moment:
(a) AV14, 8 channel $\langle \, \mbox{$^{3}$H} \mid \mbox{$^{3}$H} \rangle$ 
(b) AV14, 8 channel $\langle \, \mbox{$^{3}$H} \mid \mbox{$^{3}$He} \rangle$   
(c) AV14+3BF, 8 channel $\langle \, \mbox{$^{3}$H} \mid \mbox{$^{3}$He} 
\rangle$ 
(d) AV14+3BF, 22 channel $\langle \, \mbox{$^{3}$H} \mid \mbox{$^{3}$He}
\rangle$ }
\label{3Nmag}
\begin{tabular}{lrrrrr}
          & Ref.~\cite{LAmag} 
& \multicolumn{1}{c}{(a)} 
& \multicolumn{1}{c}{(b)} 
& \multicolumn{1}{c}{(c)} 
& \multicolumn{1}{c}{(d)}\\ \hline
IA       & --2.172 & --2.177 &--2.174& --2.175& --2.175\\
pair     & --0.290 & --0.288 &--0.285& --0.298& --0.298\\
pion     &   0.092 &   0.096 &  0.095&   0.100&   0.101\\
delta    & --0.099 & --0.100 &--0.098& --0.104& --0.105\\
total    & --2.469 & --2.468 &--2.462& --2.477& --2.477
\end{tabular}
\end{table}

\mediumtext
\begin{table}
\caption{Trion Gamow-Teller matrix element. The entries labeled 
AHHST are taken from Ref.~\protect\cite{AHHST} and use wavefunctions found 
from the Paris potential.}
\label{GT}
\begin{tabular}{ccdddd}
Current & AHHST & local & AHHST & non-local & AHHST \\ \hline
AP1+AP7         & [2a,vert] & 0.000 & 0.000 & --0.016 & --0.015 \\
AP3            & [2a,ret]  & 0.000 & 0.000 & --0.008 & --0.007 \\
AP4            & [2a,form] & 0.000 & 0.000 & --0.002 & --0.001 \\
APSPV          & [2a,PS--PV]& 0.000 & 0.000 &  0.016 &  0.015 \\
AP5            & [2d]      & 0.022 & 0.019 &  0.004 &  0.003 \\
ANP1           & [2c]      &--0.013 &--0.011 & 0.000  & --0.001 \\
ANP2           & [2b]      & 0.009 & 0.007 & 0.003 & 0.003 \\
one-body      & one-body  & 0.924 & 0.927 & 0.000 & 0.000
\end{tabular}
\end{table}

\narrowtext
\begin{table}
\caption{Ranges of parameters for the meson exchange currents.}
\label{param}
\begin{tabular}{cc}
Parameter & range \\
\hline
$\kappa_V           $ & 3.7 - 6.6       \\
$g_{\rho NN}^2/4\pi $ & 0.70 - 0.95     \\
$G_1                $ & 2.2 - 2.6       \\
$\lambda            $ &  0 - 1          \\
$f_{\pi NN}^2/4\pi  $ & 0.075 - 0.081   \\
$\Lambda_\pi        $ & 1.0 - 1.5 GeV   \\
$f_{\pi N\Delta}^2/4\pi $ & 0.23 - 0.36 
\end{tabular}
\end{table}
\widetext

\mediumtext
\begin{table}
\caption{Strong form factor fit for two values of rho anomalous coupling
\protect$\kappa_V\protect$. The values of 
\protect$f_{\pi NN}^2/4\pi\protect$ and 
\protect$g_{\rho NN}^2/4\pi\protect$ are 0.081 and 0.95 respectively. 
The formfactors are given in GeV.}
\label{OBEPfit}
\begin{tabular}{ccccc}
 & \multicolumn{2}{c}{$\kappa_V=3.7$} & \multicolumn{2}{c}{$\kappa_V=6.6$}\\
$\Lambda_\pi$ &$ \Lambda_\rho$ & 
\multicolumn{1}{c}{$\Lambda_a$ } &$ \Lambda_\rho$ &$ \Lambda_a $\\
\hline
1.0 & 1.86 & 1.09 - 1.86 & 1.18 & 1.09 - 1.18 \\
1.2 & 2.23 & 1.09 - 2.23 & 1.25 & 1.09 - 1.25 \\
1.5 & 2.76 & 1.09 - 2.76 & 1.33 & 1.09 - 1.33 
\end{tabular}
\end{table}

\mediumtext
\begin{table}
\caption{Results for the rate and spin observables.}
\label{res1}
\begin{tabular}{crddr}
& \multicolumn{1}{c}{AV14+3BF} & \multicolumn{1}{c}{AV14+3BF} &
\multicolumn{1}{c}{AV14} & \multicolumn{1}{c}{EPM} \\ &
\multicolumn{1}{c}{[22]} & \multicolumn{1}{c}{[8]} &
\multicolumn{1}{c}{[8]} & \multicolumn{1}{c}{Ref.\ \cite{JCHF}} \\
\hline
$\Gamma_0$ [s$^{-1}$] & \multicolumn{1}{c}{1502 $\pm$ 32} & 
\multicolumn{1}{c}{1498} & \multicolumn{1}{c}{1456} & 
\multicolumn{1}{c}{1497 $\pm$ 21} \\
$A_v     $ &  0.515 $\pm$ 0.005 &  0.515  &  0.516 & 0.524 $\pm$ 0.006 \\
$A_t     $ &--0.375 $\pm$ 0.004 &--0.375  &--0.373 &--0.379 $\pm$ 0.004 \\
$A_\Delta$ &--0.110 $\pm$ 0.006 &--0.110  &--0.110 &--0.097 $\pm$ 0.007
\end{tabular}
\end{table}

\mediumtext
\begin{table}
\caption{Results for the rate and spin observables. The one-body
 currents are taken from the 22 channel AV14+3BF wavefunctions.}
\label{res4}
\begin{tabular}{crdd}
& \multicolumn{1}{c}{AV14+3BF} & \multicolumn{1}{c}{AV14+3BF} &
\multicolumn{1}{c}{AV14} \\ &
\multicolumn{1}{c}{[22]} & \multicolumn{1}{c}{[8]} &
\multicolumn{1}{c}{[8]} \\
\hline
$\Gamma_0$ [s$^{-1}$] & \multicolumn{1}{c}{1502} & 
\multicolumn{1}{c}{1501} & \multicolumn{1}{c}{1491} \\
$A_v     $ &  0.515  &  0.516  &  0.518 \\
$A_t     $ &--0.375  &--0.375  &--0.374 \\
$A_\Delta$ &--0.110  &--0.110  &--0.108 
\end{tabular}
\end{table}

\widetext
\begin{table}
\caption{Contributions of each current $(\times 10^3$) to the current
 amplitudes and effective form factors.}
\label{res2}
\begin{tabular}{crrrrrrrrr}
Current & $j_\sigma$ & $j_x^{(2)}$ & $\rho_1$ & 
          $j_Q$      & $j_x^{(1)}$ & $\rho_0$ & 
          $G_V$ & $G_A$ & $G_P$ \\ \hline
IA          & 877&--326&   10&15&200& 820 & 835 & 1185 & 516 \\
AP1+AP7     &--17&     &     &  &   &  &    &--17 &  \\
AP2+AP6     & --6& --19&     &  &   &  &    &     & 19 \\
 AP3        & --8&     &     &  &   &  &    &--8  &  \\
 AP4        & --3&     &     &  &   &  &    &--3  &  \\
 APSPV      &   8&     &     &  &   &  &    &  8  &  \\
 AP5        &  22&     &  1.1&  &   &  &    & 22  &--1\\
 ANP1       &--13&     &--0.1&  &   &  &    &--13 &  \\
 ANP2       &   8&     &  0.4&  &   &  &    &  8  &  \\
 ANP3       & --2& --5 &     &  &   &  &    &     & 5 \\
 ANP4       &  77&   4 &     &  &   &  &    &  75 &--4\\
 ANP5       &--29&     &     &  &   &  &    &--29 &   \\
 ANP6       &   2&     &     &  &   &  &    &   2 &   \\
 VP1        &    &     &     &  & 64&  &    &  64 & 64 \\
 VP2        &    &     &     &  &--8&  &    & --8 &--8\\
VNP1        &    &     &     &  & 16&  &    &  16 & 16 \\
VNP2        &    &     &     &  &--6&  &    & --6 & --6 \\ \hline
total       & 918&--346&  12 &15& 267&820& 835& 1300 &601\\
 EPM        & 928&--372&  11 &15& 241&839& 854& 1293 &603
\end{tabular}
\end{table}

\narrowtext
\begin{table}
\caption{The split into local and non-local contributions. 
Only those currents which receive contributions from both 
local and non-local currents are included.}
\label{res3}
\begin{tabular}{crrrr}
Current 
& \multicolumn{2}{c}{$j_\sigma (\times 10^3)$} & 
  \multicolumn{2}{c}{$j_x^{(2)}(\times 10^3)$} \\ \hline
        & local & non-local & local & non-local \\
AP2+AP6 & --4.2 & --2.1 & --12.6 & --6.4 \\
AP5     &  18.9 &   3.1 & \\
ANP1    &--12.8 &   0.2 \\
ANP2    &   6.2 &   2.2 \\
ANP3    & --1.2 & --0.4 & --3.6 & --1.1 
\end{tabular}
\end{table}

\narrowtext
\begin{table}
\caption{Sensitivity of observables ${\cal O}$ to $g_P$.}
\label{res5}
\begin{tabular}{cd}
${\cal O}$ & $\left | \frac{g_P}{\cal O} \frac{d {\cal O}}{d g_P}
\right | _{g_P^{\rm PCAC}} $ \\
\hline
$\Gamma_0$ & 0.11 \\
     $A_v$ & 0.37 \\
     $A_t$ & 0.73 \\
$A_\Delta$ & 0.75
\end{tabular}
\end{table}

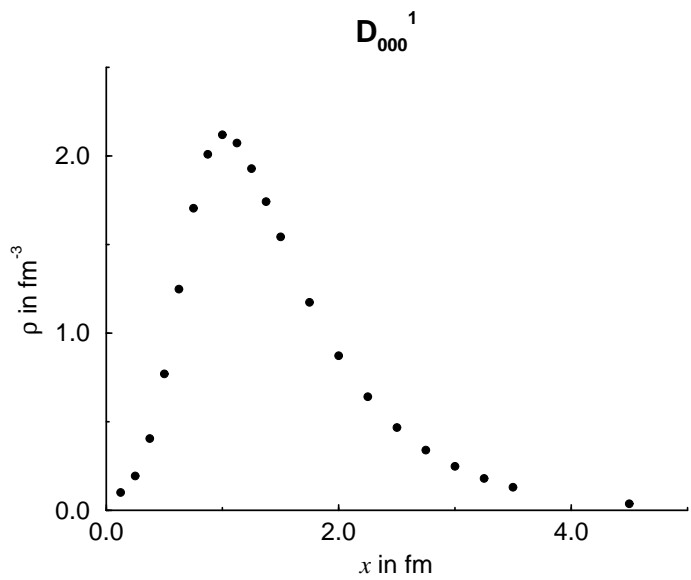
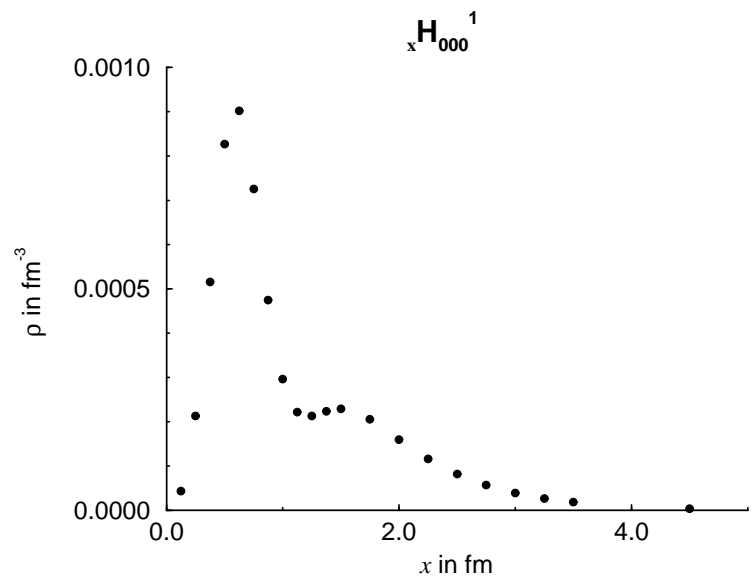
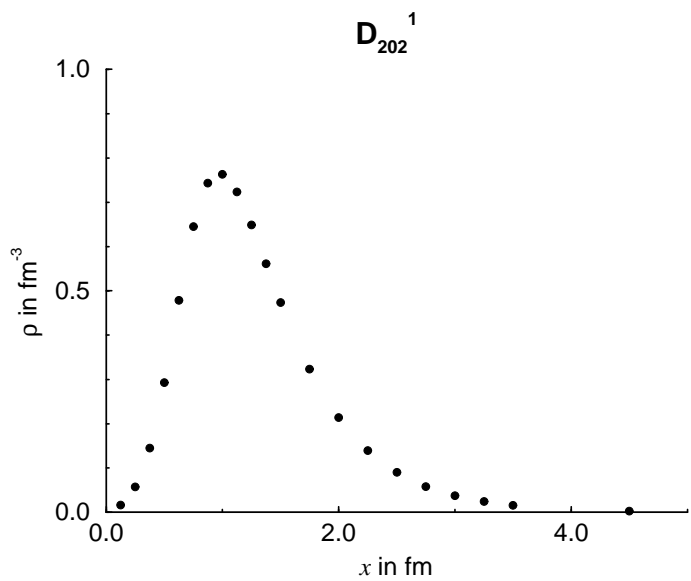
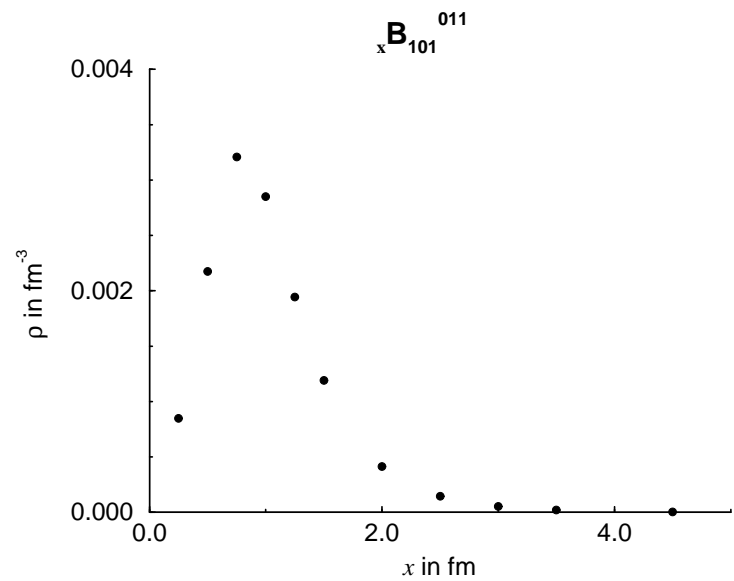
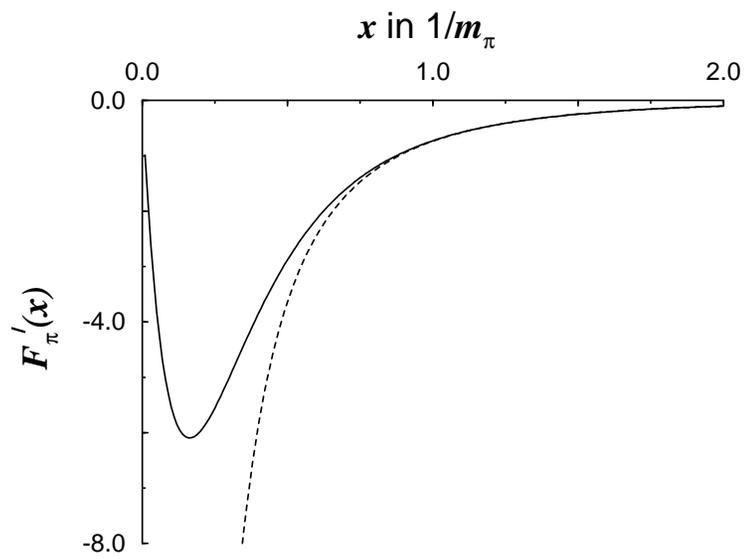
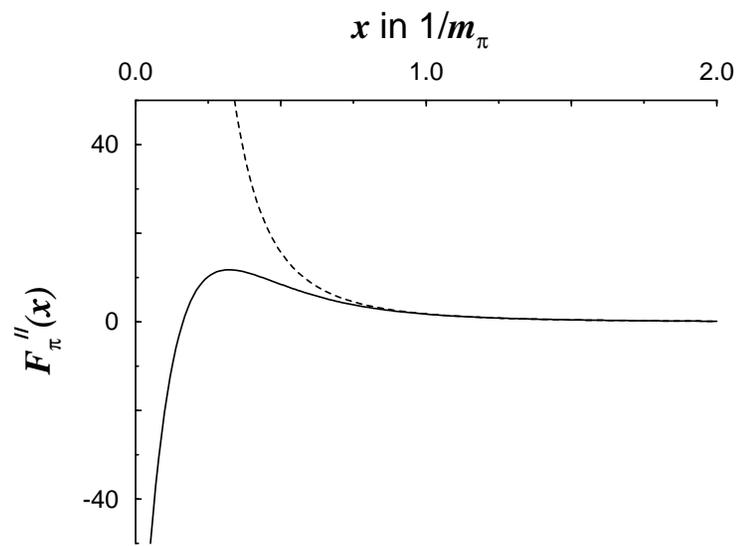
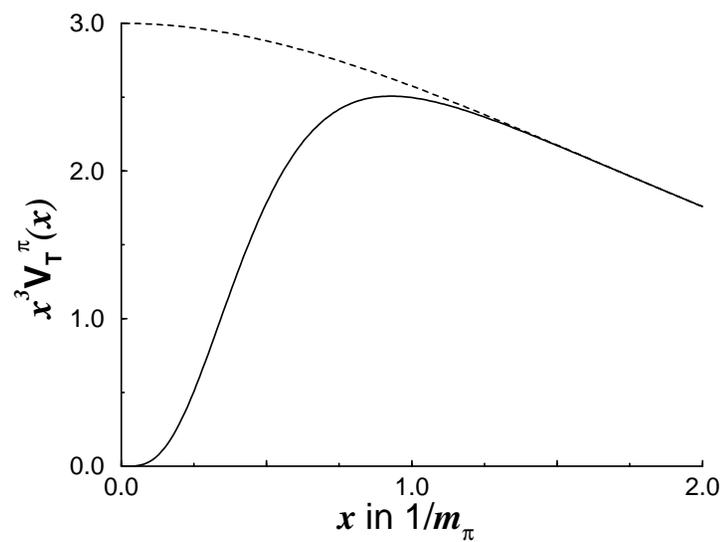
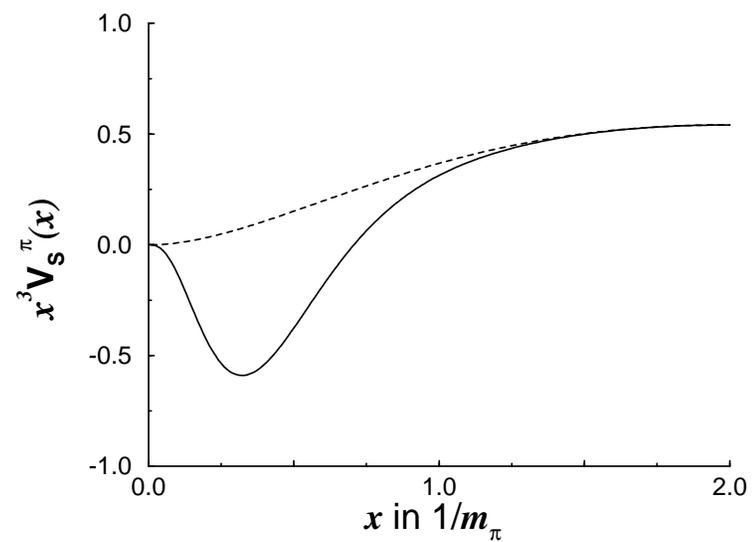
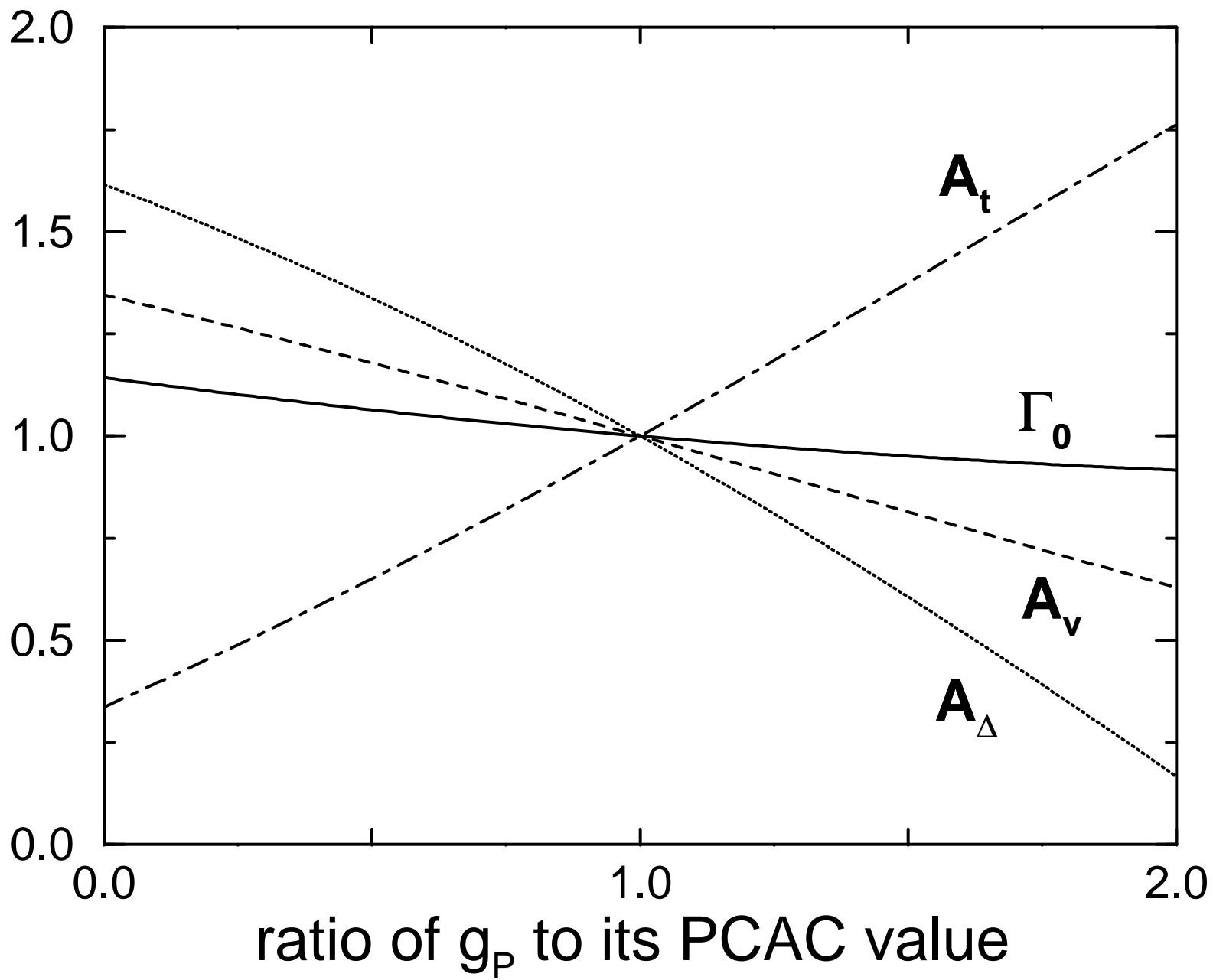
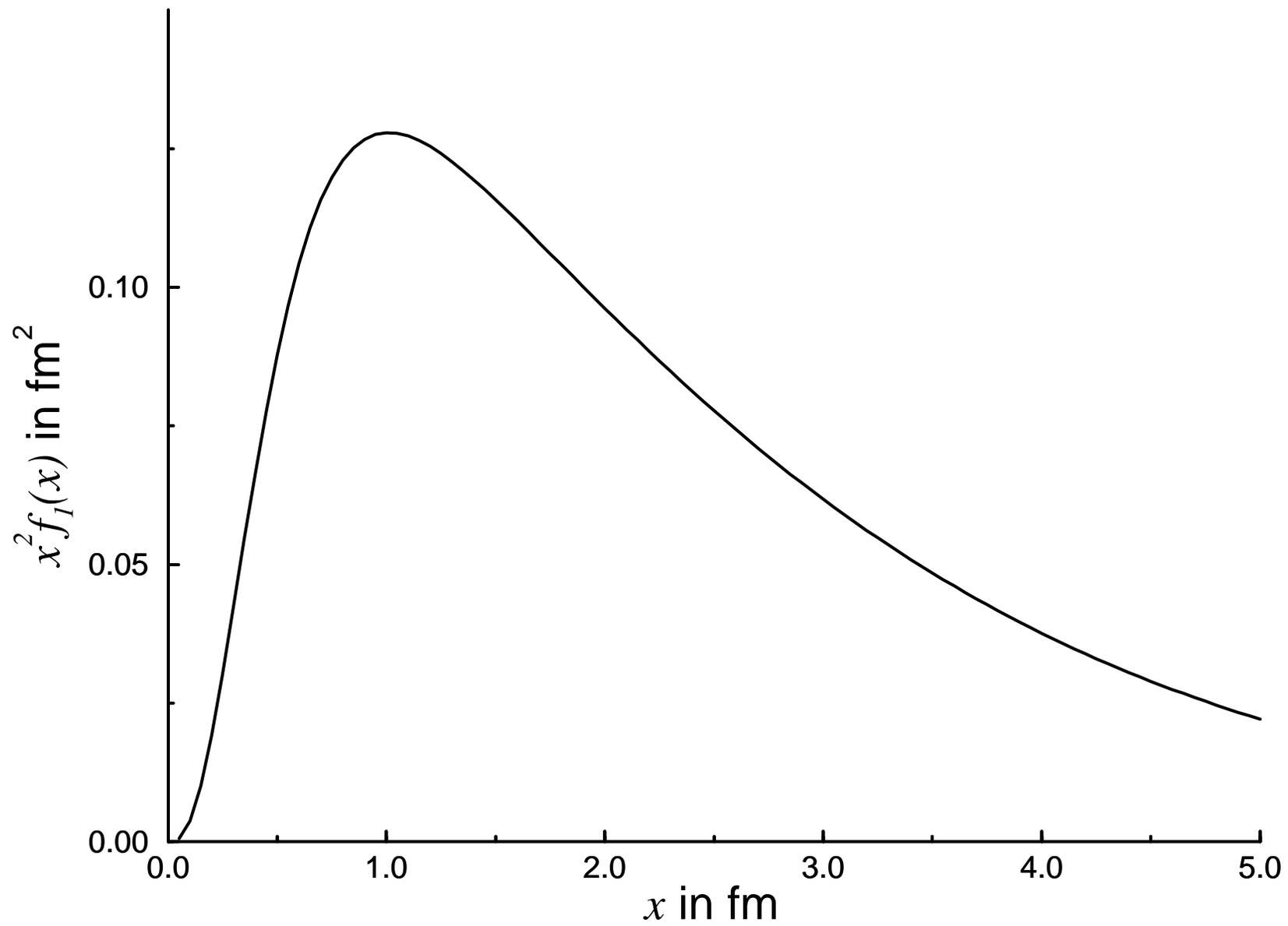
\end{document}